# DNA unzipping phase diagram
# calculated via replica theory

C. Brian Roland
Chemical Physics Program, Harvard University, Cambridge, MA 02138

Kristi Adamson Hatch
Department of Physics, Harvard University, Cambridge, MA 02138

Mara Prentiss
Department of Physics, Harvard University, Cambridge, MA 02138

Eugene I. Shakhnovich
Department of Chemistry and Chemical Biology, Harvard University,
Cambridge, MA 02138

We show how single-molecule unzipping experiments can provide strong evidence that the zero-force melting transition of long molecules of natural dsDNA should be classified as a phase transition of the higher-order type (continuous). Towards this end, we study a statistical mechanics model for the fluctuating structure of a long molecule of dsDNA, and compute the equilibrium phase diagram for the experiment in which the molecule is unzipped under applied force. We consider a perfect-matching dsDNA model, in which the loops are volume-excluding chains with arbitrary loop exponent $c$. We include stacking interactions, hydrogen bonds, and main-chain entropy. We include sequence heterogeneity at the level of random sequences; in particular, there is no correlation in the base-pairing energy from one sequence-position to the next. We present heuristic arguments to demonstrate that the low-temperature macrostate does not exhibit degenerate ergodicity-breaking. We use this claim to understand the results of our replica-theoretic calculation of the equilibrium properties of the system. As a function of temperature, we obtain the minimal force at which the molecule separates completely. This critical force curve is a line in the temperature-force phase diagram that marks the regions where the molecule exists primarily as a helix, versus the region where the molecule exists as two separate strands. We compare our random-sequence model to magnetic tweezer experiments performed on the 48502 bp genome of bacteriophage λ. We find good agreement with the experimental data, which is restricted to temperatures between 24 and $50°C$. At higher temperatures, the critical force curve of our random-sequence model is very different for that of the homogeneous-sequence version of our model. For both sequence models, the critical force falls to zero at the melting temperature $T_c$ like $|T - T_c|^\alpha$. For the homogeneous-sequence model, $\alpha = 1/2$ almost exactly, while for the random-sequence model, $\alpha \approx 0.9$. Importantly, the shape of the critical force curve is connected, via our theory, to the manner in which the helix fraction falls to zero at $T_c$. The helix fraction is the property that is used to classify the melting transition as a type of phase transition. In our calculation, the shape of the critical force curve holds strong evidence that the zero-force melting transition of long natural dsDNA should be classified as a higher-order (continuous) phase transition. Specifically, the order is 3rd or greater.

## I. Introduction

The metabolism of genomic DNA – replication, transcription, and recombination – are processes that are fundamental to the reproduction, function, and evolution of organisms. To understand the action of the proteins that carry out genomic DNA metabolism, it is necessary to know the structure assumed by double-stranded DNA (dsDNA) in a cell, as well as the amount of energy required to change that structure. Despite the simplicity of Watson and Crick's double-helical structure of crystallized dsDNA [1], when long dsDNA is in contact with a constant-temperature fluid, as is genomic DNA in a cell, spontaneous thermal fluctuations can cause a small region of the double helix to unwind and separate (denature). Clearly, the prevalence of locally denatured regions bears on questions regarding the mechanisms by which proteins achieve the metabolism of the genome. Thus, since the 1950's, it has been the objective of both experimental and theoretical endeavors to understand how the physical structure of DNA constrains the space of possible biological activities. One step towards this objective is to understand the manner in which dsDNA's physical structure changes as temperature is varied.

A standard practice for determining the prevalence of locally denatured regions (loops) is by spectroscopically measuring the helical content of a solution of DNA molecules, and comparing the temperature dependence of the reading with the helix fraction (fraction of base-pairs in involved in complementary base-pairing) predicted by models. There is evidence from bulk spectroscopy measurements, that as the temperature is increased over an interval of $10-15°C$, long dsDNA melts in a sequence of local unwinding events [2,3]. But recently, new experimental methods have been developed that complement the information obtained from spectroscopic helicity studies.

Single-molecule force experiments have been performed in which a polynucleotide double-helix is "unzipped". In this experimental configuration, a dsDNA molecule or an RNA hairpin is pulled apart by an apparatus that exerts a constant force that tends to separate the base-pair at one end of the double-helix [4,5]. Danilowicz et al. experimentally determined the minimal force, termed the critical force, to completely separate the strands of a 1500 bp sub-sequence of the 48502 bp genome of the bacteriophage λ. The critical force is essentially a measure of the stability of the duplex state as compared to the strand-separated state. This critical force was determined as a function of temperature, and the resulting critical force curve is a line in a temperature-force phase diagram that marks the regions where the molecule exists primarily as a helix, versus the region where the molecule exists as two separate strands. We study a model for unzipping the entire genome of phage λ (not just the 1500 bp sub-sequence), and show how the measurement of the critical force curve in the neighborhood of the melting temperature will allow us to classify the genome's melting transition as a type of phase transition.

In the statistical mechanical theory of macroscopic systems, an extended version [6] of the Ehrenfest classification scheme [7] can be used to describe the manner in which a macroscopic observable changes abruptly as a macroscopically controllable parameter is varied. The partition function $Z(T)$ of the system is a function of the macro-parameters (among them is temperature $T$) and thus so is the free energy density $f(T) = -RT \ln Z(T)/N$; R is the universal gas constant and $N$ is proportional to the number of particles in the system. In the extended Ehrenfest scheme, we say that there is a $kth$-order phase transition at temperature $T_{tr}$ if a $kth$ derivative of the free energy density with respect to any combination of the macro-parameters is discontinuous (has a jump, which is a discontinuity of the simple kind, or a divergence, which is a discontinuity of the non-simple kind[8]) but all derivatives of lower order are continuous (smooth). In this article, the term higher order means that the phase transition is not 1st order. Additionally, we might specify the kind of discontinuity with the terms jump kth order or divergent kth order. In general, 1st-order transitions are jump and high-order transitions are divergent, so unless otherwise indicated, this is to be understood.

In a fundamental work towards the classification of the melting transition of long dsDNA, Poland and Scheraga considered the perfect-matching (PM) model – loops are allowed, but base-pairing can only occur between complementary bp's – with a homogeneous sequence [9]. A loop of length $\ell$ has a free energy that increases (destabilizes) like $c \ln \ell$, resulting in a "long-range interaction" between the helix-state bp's flanking the loop; $c$ is a parameter termed the loop exponent that depends upon the model for the structure of a loop. Despite the approximate one-dimensionality of dsDNA, they showed that the long-range interaction causes the melting transition to be a true phase transition, and the order depends upon the parameter $c$. Because this work used analytical methods that employed the infinite-chain limit, it produced precise and succinct statements about the order of the phase transition (a phase transition is defined only for infinite systems[10]), in the homogeneous sequence approximation. But, the homogeneous PM model fails to describe long molecules of natural dsDNA, e.g. the genome of phage λ, because the width of the model's transition is much smaller than observed in spectroscopic studies [10]. Thus, we turn to models in which sequence heterogeneity is incorporated with random sequences.

Our first question is whether or not the long-range interaction present in the loop entropy considered by Poland and Scheraga is a necessary component of a random-sequence model. Fisher numerically simulated a model of loop structure that predicted $c \approx 1.75$, so that in our theory of real dsDNA, $c > 0$ [11]. But, there have been many investigations of random-sequence 1-dimensional nearest-neighbor (NN) Ising models for dsDNA – including the exact treatment of Lehman and McTague [12], and the approximate treatments of Lifson [13], Lifson and Allegra [14], and Poland and Scheraga [15] – which can be considered as approximations to the PM model with $c = 0$. But every heterogeneous 1-dimensional NN Ising model is less cooperative than the homogeneous version, which does not have a true phase transition [16], even in the long-chain limit. Thus every heterogeneous 1-d NN model differs qualitatively from the homogeneous PM model. So, we cannot say, a priori, whether or not a random-sequence NN Ising models will be a safe approximation to the random PM model, which includes the physics of $c > 0$. Therefore, as a model for the 48502 bp genome of phage λ, we choose to study the random-sequence PM model in the long-chain limit, and attempt to determine if this model melts with a true phase transition, and thus resembles more closely the homogeneous PM model or the random NN Ising model. If the random PM model does have a true phase transition, we attempt to classify it by the order.

Our model includes (a) perfect-matching (b) uncorrelated random sequence of base-pairing free energies (c) loops with arbitrary loop exponent $c$ (d) end sequence that couples to external unzipping force (e) long-chain limit. We give a heuristic argument that the low-temperature macrostate does not exhibit degenerate ergodicity-breaking. We use the no ergodicity-breaking statement to interpret the results of our replica calculation, which give the model's thermodynamics for all experimentally relevant temperatures. We find that the presence of sequence heterogeneity changes the order of the melting transition from $1^{st}$-order to higher-order (continuous).

The organization of the paper is as follows. In Section II, we describe a microscopic model and a useful reduction in description. In Section III, we present a version of the replica method that is amenable to our model. In Section IV, we describe the zero-force thermodynamics resulting from our approximations, showing that the sequence heterogeneity changes the order of the melting transition from $1^{st}$- to higher-order, when the loop exponent $c = 2.115$. We present the temperature-force phase diagram, and in Section V, we show how the shape of the critical force curve near the melting temperature is determined by an effective value of the loop exponent $c^{ef}$, which evidences the order of the melting transition as $3^{rd}$ or greater.

## II. The model

### A. Mechanical model in terms of sequence partition functions

#### 1. Sequence partition functions

In our minimal model of dsDNA unzipping, we retain only those features of molecular structure necessary to analyze the interplay between base-pairing heterogeneity and thermally induced local melting (loops), and whether this interplay has consequences for the temperature-force phase diagram. Firstly, we choose a partitioning of the strands into "nucleotides" that is convenient to describe the physics of unzipping, as shown in Figure 1(a). We label the "nucleotides" with sequence-position $t$, running from 1 to N. Next, we assume that each complementary base-pair (bp) may exist in one of two microstates: the helix state or the coil state. The helix state of bp $t$ is defined by the activation of the complementary hydrogen bonds (h-bonds), and to achieve this, the bases are close and properly oriented [see Figure 1(b)]. We assume that if bp's $t$ and $t-1$ are both in the helix-state, then for each strand, the main-chain bonds linking the deoxyribose of bp $t$ to the deoxyribose of bp $t-1$ are frozen into a special rotation state. We ignore the rotation states provided by the bond connecting deoxyribose to the nitrogenous base, and lump this entropy into that of the main-chain bonds. The coil state of bp $t$ is defined by the deactivation of the h-bond, and thus the bases are far apart or mis-oriented. Thus, if bp $t$ or $t-1$ are in the coil state, then the main-chain bonds of bp $t$ are free to sample many rotation states. In the experiment[4], a hairpin is synthetically added to the not-unzipped end of the double-helix; this is modeled by imagining a fictitious bp at sequence-position $t=0$ that remains forever in the helix state.

From the definitions of single bp microstates, we may state our algorithm for assembling the statistical weight to observe a particular microstate of the entire molecule, when at equilibrium. To do this, we apply a physical approximation that is common in the statistical mechanics of polynucleic acids[10]: in any given microstate of the molecule, the only bp's that interact with a given bp are those in its mechanical sequence.

We define two types of mechanical sequences. A *bound sequence* is a contiguous sub-chain of bp's such that all but the last (according to sequence-position $t$) are in the coil state; the last bp is in the helix state [see Figure 1(b)]. The bp preceding (in sequence-position) the bound sequence is in the helix state, and a bound sequence of length one is a single helix-state bp. For any given microstate of the model, all bps up to and including the last helix-state bp may be assigned to a bound sequence. All bps following the last helix-state bp belong to the second type of mechanical sequence, which is called the *unbound sequence*. All bp's are in the coil state, and experience the stability offered by the external force $F$. In the statistical mechanics of polynucleic acids, there are two distinct schemes for partitioning the bound section of the model into mechanical sequences. We follow the scheme of Gibbs and DiMarzio[17], where the bound section of the model is made up of a series of bound sequences. Alternatively, the bound section can be described as an alternating series of *helix* and *loop* sequences, as in the manner of Hill[18]. For mathematical convenience, we use the scheme of Gibbs and DiMarzio.

According to our physical approximation, the statistical weight to observe any given polymer microstate is a product of the statistical weight factors (partition functions) for each mechanical sequence present in that microstate. The statistical weight factor (partition function) for a bound sequence of length one (a single helix-state bp preceded by a helix-state bp) is

$$e^{-\frac{g_t}{RT}} \quad (1)$$

where

$$g_t = h - Ts + \delta g_t \quad (2)$$

is called the *helix-propagation free energy,* and is the energy to transfer a bp at sequence-position $t$ from the coil to the helix state, given that bp $t-1$ is in the helix state [see Figure 1(b)]. The average of $g_t$, over all bp's in the entire molecule is

$$h - Ts = \frac{1}{N} \sum_{t=1}^{N} g_t \quad (3)$$

where $h$ and $s$ are the enthalpic and entropic components of this average. We associate $h$ with the stacking interaction between bp's $t$ and $t-1$, and associate $s$ with the main-chain bonds of bp $t$ [see Figure 1(b)]. For real dsDNA, the stacking interaction will depend upon the chemical identities of bp $t$ and bp $t-1$. But here we assume that the stacking interaction is homogeneous, and lump the sequence dependence of the interaction into the fluctuation term $\delta g_t$, which in our model is associated only with the hydrogen-bonds between the bases of bp $t$.

The statistical weight factor (partition function) for a bound sequence with $\ell \geq 2$ bp's, terminated at sequence-position $t$, is

$$\frac{A}{\ell^c} e^{-\frac{\delta g_t}{RT}} \quad (4)$$

where the parameters have the following thermodynamic meaning. Suppose we have some sequence of $\ell$ coil-state bp's, preceded by a bp in the helix state, and followed by no other base-pairs (see Figure 2). The macrostate of the sequence – the free coil macrostate – has a T-independent free energy (per bp) that we set to zero, i.e., we measure all free energies with respect to the free coil macrostate (here, we imagine $\ell \to \infty$). If we constrain the terminal bp in the sequence to be in the helix state, we have formed a loop and this action will raise the free energy of the sequence above the zero of free energy. The factor $A/\mathbb{1}^c$ gives the entropic cost to limit the rotation states of the "nucleotides" of the strands, so that they may form a loop. The factor $\exp[-\delta g_t/RT]$ represents the free energetic gain/loss upon transferring the terminal bp from the coil to the helix state, given that bp $t-1$ is in the coil state (the h-bond is activated).

The T-independent prefactor $A$ is associated with the configuration-space volume of the helix-state of the terminal bp. This configuration-space volume is achieved by restricting the main-chain bonds of all "nucleotides" in the bound sequence but that the associated entropy is numerically different from the entropy $s$ in Eq. (3), as there is no helix-state bp preceding the terminal bp. We have chosen the notation $A$ because it emphasizes that it is purely entropic in origin, and because it is consistent with [9, 19, 20]. The parameter $A$ is given the name (constant) in section 9D of [10]. Because $A$ plays a role similar to the cooperativity parameter $\sigma$ in models of polyalanine[21], it is often called $\sigma_c$, as in [22], or $\sigma$ as in [23]; note that in polyalanine models, $\sigma$ has an enthalpic origin.

The parameter $c$ governs the manner in which the number of configurations – relative to the coil state – of a loop decreases as the length of the loop increases. Fisher showed that if we model the loop as a volume-excluding polymer, but one that is isolated from other portions of the chain, then the appropriate value is $c = 1.75$ [11]. Kafri et al. showed that if we include interactions with the other portions of the chain outside the loop, the appropriate value is at least $c = 2.115$ [19]. Recently, Blossey et al. remarked that the Kafri model may be inappropriate for real dsDNA because the difference in the persistence lengths of double-stranded (in helix regions) and single-stranded DNA (in loops) was ignored [23]. For this reason, we consider both the Fisher and Kafri values, $c = 1.75$ and $c = 2.115$, respectively. Employing the numerical algorithm MELTSIM of Blake et al. [24], Blossey et al. fit the parameter $A$ (in those works it is called $\sigma$) to each value of $c$: for $c = 2.115$, $A = 1.75 \times 10^{-4}$, and for $c = 1.75$, $A = 1.26 \times 10^{-5}$.

The statistical factor (partition function) for an unbound sequence of $\ell_F$ bp's is

$$e^{-\frac{g_F}{RT}\mathbb{1}_F} \tag{5}$$

and $g_F$ is the free energy per bp in this sequence. The statistical factor in Eq. (5) factorizes (exactly) by bp because we model the unbound sequence as a freely-jointed-chain under stretching force [25], obtaining

$$g_F(T,F) = -\frac{a}{b} RT \log\left[\frac{k_B T}{Fb} \sinh\left(\frac{Fb}{k_B T}\right)\right] \tag{6}$$

where $F$ is the applied force in pN, $T$ is the temperature in Kelvins, $b = 1.9$ nm is the Kuhn length of ssDNA in the solvent conditions of the experiment. Because $g_F$ is the free energy per bp, we include the prefactor $a/b$, where $a = 0.7$ nm is the length of a "nucleotide" in ssDNA. The parameter $k_B$ is Boltzmann's constant, and the appearance of the gas constant $R$ means that $g_F$ is in units of per mol of bp's.

## *2. Polymer in 1 dimension*

With the sequence partition functions described above, we can give the statistical factor for any given microstate. But it proves convenient to describe the equilibrium statistics of our mechanical model in term of the following hamiltonian

$$H_g[\rho] = \sum_{i=1}^{N} u_A(\rho_i, \rho_{i-i}) + \sum_{i=1}^{N} u_g(\rho_i, \rho_{i-i}) + \sum_{i=1}^{N} \delta g_{\rho_i} + g_F \cdot (N - \rho_N) \qquad (7)$$

which describes the fictitious 1-dimensional polymer shown in Figure 3; Eq. (7) can be derived from a the sequence partition function description. The polymer is composed of monomers that represent the helix-state bp's of the dsDNA model. The monomers of the 1-d polymer are labeled $i \in \{0, K, N\}$, each having position variable $\rho_i \in \{0, K, N\}$. Degree of freedom $\rho_0$ is fixed at $0$, while for $i \in \{1, K, N\}$, $\rho_i$ is a fluctuating variable (of the annealing type). Every monomer $i$ is confined to the position-space $\{0, K \ N\}$ by walls at positions $0$ and $N$. The initial configuration of the polymer is any arrangement satisfying $\rho_{i-1} \leq \rho_i$ for each pair of nearest-neighbor monomers.

The polymer connectivity is described with nearest-neighbor pair interaction

$$u_A(\rho_i, \rho_{i-1}) = \begin{cases} 0 & , \rho_i = 0 \\ \Lambda T & , \rho_i > 0 \text{ and } \rho_i - \rho_{i-1} = 0 \\ 0 & , \rho_i > 0 \text{ and } \rho_i - \rho_{i-1} = 1 \\ T \log A^{-1} + cT \log(\rho_i - \rho_{i-1}) & , \rho_i > 0 \text{ and } \rho_i - \rho_{i-1} \geq 2 \end{cases} \qquad (8)$$

where the first line of Eq. (8) indicates that when a monomer is at $t = 0$, the interaction with the previous monomer is turned off. When a monomer is at any $t > 0$, it experiences a strong repulsion with the previous monomer at short separations and a weak attraction at large separations (see Figure 4). The parameter $\Lambda$ is a large number that makes it very unfavorable to find two monomers at the same position. As related to the dsDNA model, the potential $u_A$ represents the effective interaction between the helix-state bp's terminating and preceding a bound sequence and is due to the entropy of the intervening coil-state bp's.

Two monomers that are next to each other in position-space share an interaction

$$u_g(\rho,\rho') = \begin{cases} \text{h-Ts}, & \rho,\rho' > 0 \text{ and } |\rho - \rho'| = 1 \\ 0, & \text{otherwise} \end{cases} \qquad (9)$$

This short-range attractive potential has the same form as that in the 1-dimensional lattice-gas [26]. Because monomers cannot move past one another, only nearest-neighbors along the chain may share this interaction. As related to the dsDNA model, the interaction $u_g$ corresponds to the stacking and main-chain entropy contributions to the helix-propagation free energy. Note that we ignore the helix-propagation energy between monomer $i$ at $t = 1$ and monomer $i - 1$ at $t = 0$, but this interaction is negligible because it makes a non-thermodynamic contribution to the hamiltonian [see Eq. (7)].

The monomers interact with an external potential field $\delta g_t$ (sequence-dependent h-bonding in dsDNA model) that depends on position $t$. Additionally, monomer $N$ is driven to the left by compressing force $g_F$, which represents the interaction between the unbound sequence and the unzipping force in the dsDNA model [see Figure 3].

For this model, the partition function is

$$Z_g(T,F) = Tr_\rho e^{-\frac{H_g[\rho]}{RT}} \qquad (10)$$

where the notation $Tr$ represents the sum over all arrangements of the monomers that preserves $\rho_{i-1} \leq \rho_i$ for each pair of nearest-neighbors.

**B. Genetic model**

*1. Introduction*

In addition to the above mechanical model, we must explain our model for the genetic sequence of λ-phage DNA, i.e., we must specify the helix propagation free energies $g_t$. We assume that thermodynamic observables of interest can be well approximated by modeling the genetic sequence of λ-phage dsDNA as random. Specifically, we consider a large ensemble of sequences, in which each sequence is generated by choosing each $g_t$ from a gaussian distribution with mean $g^{(1)}$ and standard deviation $g^{(2)}$. As a result of training these parameters on experiment – described next – $g^{(1)}$ and $g^{(2)}$ depend on temperature $T$ and sodium concentration $[Na^+]$, but we do not explicitly indicate these dependencies in our notation.

For convenience, our mechanical model is written in terms of *helix-propagation* parameters $g_t$, representing the free energy to constrain bp $t$ to the helix state, given that bp $t-1$ is in the helix-state. For each sequence in the sequence-ensemble, if each bp is constrained to the helix-state, the molecule will have a free energy (per bp)

$$g^{(1)} = \frac{1}{N} \sum_{t=1}^{N} g_t \tag{11}$$

and the variance (from one sequence-position to the next) in free energy (per bp) is

$$[g^{(2)}]^2 = \frac{1}{N} \sum_{t=1}^{N} g_t^2 - [g^{(1)}]^2 ; \tag{12}$$

fluctuations in $\sum g_t$ decay like $1/N^{1/2}$. The random variables $g_t$ can be connected to the DNA melting literature by fixing the sequence-ensemble parameters to the statistical parameters of the genome, i.e., $g^{(1)} = g_\lambda^{(1)}(T)$, $g^{(2)} = g_\lambda^{(2)}(T)$ where

$$\begin{aligned}
g_\lambda^{(1)}(T) &\equiv \frac{1}{N} \sum_{t=2}^{48502} g^{Blake}(b_{t-1}, b_t; T) \\
[g_\lambda^{(2)}(T)]^2 &\equiv \frac{1}{N} \sum_{t=2}^{48502} g^{Blake}(b_{t-1}, b_t; T)^2 - [g_\lambda^{(1)}(T)]^2
\end{aligned} \tag{13}$$

where in Eq. (13) and (14) the index $t$ runs over the genome of λ-phage, beginning at the "left" end of the "l-strand"; the l-strand has a $5'$ G, and its left end is the $5'$ end, which turns out to have a higher $GC$-content than the right end (notation from [27]). The parameter $g^{Blake}(b_{t-1}, b_t; T)$ is the contribution due to the dimer $5' - b_{t-1}b_t - 3'$, where $b_t$ is the letter of the base at sequence-position $t$, to the total free energy to transfer the genome from the free coil (strands separated) state to the helix macrostate (each bp in the helix state.) The parameters $g^{Blake}(b, b'; T)$, one for each type of NN dimer, were experimentally determined in the experimental study of Blake and Delcourt [22], by fitting the nearest-neighbor (NN) model of Tinoco and Uhlenbeck[28] to spectroscopic melting measurements. We choose the parameters obtained in the Blake-Delcourt study, as opposed to other studies, because the size of the molecules used in this experiment most closely resembles that of the 48502 bp genome of λ-phage. To the entropic component of the Blake-Delcourt parameters, we add a salt dependence using column 2 in Table 2a of [22]. In this paper, we fix $[Na^+] = 0.14$ $M$, as this approximates the ionic condition of phosphate-buffered saline (PBS), the solvent used in the unzipping experiments [4].

Later, we will compute the thermodynamics of each sequence in the ensemble, and average the result over all members in the ensemble; these procedures are integrated into a single analytical process. We assume that the thermodynamics of each sequence self-averages (see [25, 29] for discussions of self-averageness), so that the sequence-ensemble average is a good proxy for the thermodynamics of any given random sequence.

Note that in our statistical ensemble of sequences, each sequence exhibits zero correlation between $g_t$ and $g_{t-1}$. This is an approximation because such correlation is clearly present in λ-phage dsDNA. Suppose that the base at position $t$ of the l-strand is $A$. The NN interaction parameter between bp $t$ and $t-1$ will likely be weak, and the interaction parameter for bp $t+1$ and $t$ will also likely be weak. Such correlations are ignored.

### *2. Analysis of the lambda phage genome*

We downloaded the sequence of the enterobacteria phage λ genome (NC 001416) from http://www.ncbi.nlm.nih.gov, and computed $g_\lambda^{(1)}$ and $g_\lambda^{(2)}$ as functions of $T$ (Figure 5). The components of $g_\lambda^{(1)}(T)$ are $h = -9.404$ kcal/mol and $s = -25.87$ cal/mol K.

### **C. A course-graining procedure that incorporates loops**

To study the long length-scale properties of our model near the forced unzipping transition, one may imagine course-graining the description of the DNA molecule in Section II.A, or equivalently of the 1-d polymer description [see Eq. (7)]. We first reduce the description. At a temperature where the bound macrostate is stable, i.e. $T < T_s$, the system free energy $E(k)$ is computed under the constraint that the unzipping fork, $\rho_N$, is restricted to one of a lattice of intervals of size $\ell_{course}$, labeled $k = 1,...,M$, where $k=1$ is at the end under force. The constraint is enforced by turning off the base-pairing potential for every sequence position in intervals $k' \le k$, i.e. $g^{(1)} = 0$, $\delta g_t = 0$ for $t > N - k\ell_{course}$, so that the fork naturally resides in interval $k+1$. We call $w(k)$ the change in the system free energy, at $F=0$, that results when we turn off the base-pairing potential in interval $k$ given that the potential is off in all intervals $k'$ with $k' < k$ and that the potential is on in all cells $k'$ with $k' > k$. Thus, $w(k)$ is the work to transfer interval $k$ from the bound to the free coil state at $F=0$, given that interval $k+1$ is in the bound state while interval $k-1$ is in the free coil state. Then we can write

$$E(k) = \sum_{k'=1}^{k} w(k') + k \ell_{course} g_F + O(\ell_{course}) \tag{14}$$

where the term $k \ell_{course} g_F$ suggests that we do not let the fork position advance beyond interval $k+1$ in response to the unzipping force. The term $O(\ell_{course})$, of order $\ell_{course}$, accounts for the uncertainty in the fork position within interval $k$, and thus for the uncertainty in the term $k \ell_{course} g_F$.

If we call $\ell_{seq}$ is the length-scale giving the decay of correlations in the sequence $\delta g_t$ (in our case, $\ell_{seq} = 0$), and $\ell_{loop}$ is the length-scale giving the decay of loop sizes, then the course-grained description is justified when we can choose some finite $\ell_{course}$ such that

$$\ell_{seq}, \ell_{loop} \ll \ell_{course} \ll N . \tag{15}$$

We require $\ell_{seq} \ll \ell_{course}$ so that the $w(k)$ are not correlated via correlations in the $\delta g_t$. We require $\ell_{loop} \ll \ell_{course}$ so that the $w(k)$ of adjacent intervals have correlations (due to the $\delta g_t$) that are weak in the sense that the interaction between neighboring intervals, mediated by the term $u_A$ in Eq. (8), can be made arbitrarily small relative to the magnitude of the $w(k)$. Thus, if Eq. (15) is satisfied, then the free energies $w(k)$ are uncorrelated or weakly correlated. Since we are interested in a long-length scale description, i.e. large $k$, we can utilize the central limit theorem to model the $w(k)$ as gaussian distributed and uncorrelated from each other. This model for the $w(k)$ constitutes the course-graining step in our reduction procedure.

Also, $\ell_{loop}$ gives the typical length of the free coil portion of the chain in interval

$k+1$, so $\ell_{loop} = \ell_{course}$ guarantees that the fork is indeed in interval $k+1$.

The term $O(\ell_{course})$ becomes negligible when the length of the unzipped section is much larger than $\ell_{course}$. So, in the system's thermodynamic limit, as the unzipping transition is approached and the length of the unzipped section diverges, we can ignore the $O(\ell_{course})$ term in our course-grained description, i.e. Eq. (14) with gaussian and uncorrelated $w(k)$. It is because the length of the unzipped section becomes infinite that we call the location of the forced unzipping transition $F_c$ a critical force. Note that the divergent length-scale makes possible the course-graining step.

This course-graining procedure is one way to obtain the (uncorrelated) random zipper model – a zipper model [17, 30, 31], i.e., one-sequence approximation[10], augmented with an random sequence with zero correlations amongst sequence positions – employed by Lubensky and Nelson in their study of the unzipping transition [32]. Note that we claim that the (uncorrelated) random zipper model is justified as a course-grained version of our model (see Section II.A) only when (a) the conditions in Eq. (15) are met, (b) the unzipped section becomes very large. We assume that Eq. (15) is satisfied for $T < T_c$.

The random zipper model predicts that the unzipping transition is 1st-order. Because the random zipper model is an exact reduced description of our model in the neighborhood of $F_c$, the unzipping transition of our model must be 1st order. This simplifies further analysis.

## III. Methods

### A. Sequence-ensemble averaged free energy of the bound phase

With the mechanical and genetic models in hand, we analytically compute the thermodynamics of a typical random sequence. The replica method [29] offers an analytical route to perform the sequence-ensemble averaging operation.

In the replica method, we compute the sequence-ensemble average of the free energy (per bp) as

$$-RT\frac{1}{N}\overline{\log Z_g(T)} = -RT \lim_{n \to 0} \frac{1}{nN} \log \overline{Z_g^n} \tag{16}$$

where the over-bar denotes averaging over sequences in the ensemble. We first compute, for large integer values of $n$, the sequence-ensemble average of the partition sum of $n$ uncoupled copies (replicas) of the original system to obtain:

$$\overline{Z_g^n} = Tr_{\rho^1} \ldots Tr_{\rho^n} e^{-\beta H_n[\rho^1, \ldots, \rho^n]} \tag{17}$$

where

$$\beta H_n[\rho^1, \ldots, \rho^n] = \sum_{\alpha=1}^{n} \beta H_{g=0}[\rho_\alpha] + \beta E[\langle \rho | \rho \rangle] \tag{18}$$

and $H_{g=0}$ is the hamiltonian in Eq. (7) with $g_t = 0$ for each $t$. The object

$$\beta E[\langle \rho | \rho \rangle] = -N\beta g^{(1)} \sum_{\alpha=1}^{n} \langle \rho^\alpha | \rho^\alpha \rangle - N \frac{[\beta g^{(2)}]^2}{2} \sum_{\alpha=1}^{n} \sum_{\gamma=1}^{n} \langle \rho^\alpha | \rho^\gamma \rangle \tag{19}$$

shows that, as a result of the averaging procedure, every pair of replicas $(\alpha, \gamma)$ now interacts through their overlap, defined as

$$\langle \rho^\alpha | \rho^\gamma \rangle = \begin{array}{c} \text{fraction of sequence-positions} \\ \text{exhibiting a monomer} \\ \text{in both replicas } \alpha \text{ and } \gamma \end{array} \tag{20}$$

and illustrated in Fig. 6.

The partition sum $\overline{Z_g^n}$ may be written

$$\overline{Z_g^n} = Tr_Q e^{-\beta E_n[Q] + S_n[Q]} \tag{21}$$

where

$$e^{S_n[Q]} = Tr_{\rho^1} \text{L } Tr_{\rho^n} e^{-\beta H_{g=0}[\rho^1]} \text{L } e^{-\beta H_{g=0}[\rho^n]} \delta[Q - \langle \rho | \rho \rangle] \tag{22}$$

and $\delta[Q - \langle \rho | \rho \rangle]$ constrains the $n$-replica system to a particular realization of the matrix of overlaps $\langle \rho^\alpha | \rho^\gamma \rangle$. In the factor $e^{S_n[Q]}$, we allow the n-replica system to sample all microstates consistent with $Q$.

At this point, we assume that $\overline{Z_g^n}$ is dominated by a single value of Q (maximum-term, i.e., saddle-point approximation) that is obtained by making an ansatz for the form of this matrix, and varying the corresponding parameters. The resulting $Q$ constitutes our approximate solution to Eq. (16). We employ an ansatz for $Q$ that is a restricted version of Parisi's 1-step replica-symmetry-breaking (1RSB) scheme [29]. Our RSB scheme, which we call the cuRSB, for "correlated-uncorrelated" RSB, was successfully applied in a previous theoretical study of a related random heteropolymer model [33]. We restrict the summation in Eq. (21) to a subset of the space of $Q$ matrices, and hope that the restricted sum well-approximates the unrestricted sum. Elements, $Q^c$, of the restricted space have the form shown in Figure 7 and are achieved by enforcing the cuRSB system of constraints on the n-replica system represented by $H_n$ [see Eq. (18)]: (1) the self-overlap, i.e., helix fraction, of every replica is $\theta$, (2) replicas are divided into groups of size $m$ and the overlap between any two replicas in the same group is the perfectly correlated value $q_c = \theta$, (3) the overlap between any two replicas that are in different groups is the perfectly uncorrelated value that corresponds to the situation in which the groups fluctuate independently of each other. Simply put, monomer $i$ of replica $\alpha$ and monomer $i$ of replica $\gamma$ move (a) in unison, if $\alpha$ and $\gamma$ are in the same group (b) without correlation if $\alpha$ and $\gamma$ are in different groups.

The off-block matrix elements correspond to the overlap between replicas in distinct groups. The value of the inter-group overlap $q_u$ is determined by considering the $n$-replica system $H_n$ with constraints (1) and (2) imposed and with $g^{(1)}(T) = 0$. The inter-group overlap is then the thermal average overlap between two replicas in distinct groups. It is computed as a function of $\theta$ and $m$ at $F = 0$, resulting in $q_u = \theta^2$ as shown in the Appendix A.1.

At large integer $n$, we constrain the $n$-replica system to a point in the $Q^c$ space of matrices, to obtain the free energy (per bp)

$$\beta f_n(\theta, m; T, F) = \beta g^{(1)}\theta - \frac{[\beta g^{(2)}]^2}{2}[mq_c + (n-m)q_u] - \frac{1}{mN}\log Z_{group} \qquad (23)$$

where $Z_{group}$ is the partition sum of a single group in the cuRSB constrained n-replica system, with the coupling between replicas turned off ($g^{(2)} = 0$ for all $T$). In this sense, $Z_{group}$ gives the entropic contribution to the cuRSB free energy; its explicit analytical form is given in Appendix A.1.

In the final step of this procedure, we obtain the equations that make $\beta f_n$ stationary with respect to $\theta$ and minimized with respect to $m$, solving these equations with $n = 0$. The complexities of this "minimization" are discussed in Appendix A.3.

## B. Equilibria between the bound and the unbound phase

First we imagine the system at $F = 0$ and $T < T_c$, where $T_c$ is the largest temperature below which the helix fraction is non-zero (at $F = 0$). In other words, $T_c$ is the smallest temperature above which the strands are separated, except perhaps for a thermodynamically unimportant number of helix-state bp's. We can call $T_c$ a critical temperature, in the sense of a divergent correlation length (typical loop size), if the melting transition is continuous (higher-order). For $T < T_c$, we would like to compute the value of the unzipping force $F_c$ above which the strands are separated. We call $F_c$ the critical force, because as $F \to F_c$ (from below), the system develops a divergent length-scale. Specifically, the length of the unbound section of the molecule diverges (approaches the size of the long molecule).

In order to compute the critical force $F_c$, we apply the usual maximum-term method to determine the dominant partitioning of the molecule between the bound and unbound phase. At low forces, the molecule will be dominated by the bound phase. The critical force $F_c$ is the point of phase-coexistence between the bound and unbound phases, here the free energy density of the two phases must be equal. So, at each $T$, $F_c$ is the solution of

$$f_0(\theta, m; T, F)\big|_{F=0} = g_F(T, F_c) \tag{24}$$

where $g_F$ is the free energy (per bp) of the unbound phase, given in Eq. (6).

## C. Properties of the bound macrostate

At each $T$, for $F < F_c(T)$, the majority of the molecule belongs to the bound phase, and the unbound section has an extent that is not thermodynamic (not proportional to $N$). For these values of $T$ and $F$, we say that the molecule occupies the bound macrostate. As described in Section III.A, the helix fraction $\theta$ is obtained from the replica method variational process. To describe other properties of the bound macrostate, it is helpful to imagine the bulk of the molecule as a collection of helix or internal coil (loop) sequences, in the manner of Hill [18], rather than as a collection of bound sequences. A helix sequence is a set of contiguous sequence positions at which every bp is in the helix state. A loop sequence is a string of coil-state bp's that is followed by a helix-state bp. We can compute the fraction of sequence-positions where a helix sequence is followed by a loop, i.e., by a coil-state bp. We call this fraction the density of helix-coil junctions $\theta_{hc}$, which is computed as

$$\theta_{hc} = -\partial \beta f_0(T,0)|_{F=0} / \partial \ln A \qquad (25)$$

where $\theta$ and $m$ are fixed at the values obtained in the variational process. From $\theta_{hc}$, we can compute the equilibrium average length of a helix sequence as $L_{helix} = \theta / \theta_{hc}$. Similarly, we can compute the average length of a loop sequence $L_{loop} = 1/\theta_{hc} - L_{helix}$ from the normalization condition $\theta_{hc} \cdot (L_{helix} + L_{loop}) = 1$.

## IV. Results

### A. Interpretation of the overlap matrix

Consider two uncoupled replicas of the 1-dimensional polymer system $H_g$, see Figures 3 and 6, at fixed finite N and F=0. Each replica has the same sequence $\{g_t\}$, which is picked at random from the sequence-ensemble discussed in Section II.B. The distribution of overlaps, for the sequence $\{g_t\}$, is defined as [29]

$$P_g(q) = Tr_{\rho^1} Tr_{\rho^2} \frac{e^{-\beta H_g[\rho^1] - \beta H_g[\rho^1]}}{Z_g^2} \delta\big(q - \langle \rho^1 | \rho^2 \rangle\big) \tag{26}$$

where $\delta(q - q')$ is a smoothed version of Dirac's delta and the overlap $\langle \rho^1 | \rho^2 \rangle$ is the fraction of sequence-positions at which both replicas 1 and 2 have monomers.

In Appendix A.3, we given a detailed argument that for every $T \neq T_c$, the Gibbs state of the system does not exhibit ergodicity-breaking, and thus at these temperatures, the overlap distribution $P_g(q)$ should exhibit a single peak. The location of the single peak should be a self-averaging quantity, because (1) the overlap is an intensive property of the 2-replica system and (2) we argued that the loop and tail distributions have a finite cutoff. Thus, the sequence-ensemble-averaged distribution of overlaps should exhibit a single peak. We call the distribution obtained by averaging Eq. (26) over members of the sequence-ensemble the true distribution

$$P(q) \equiv \overline{P_g(q)} = \delta(q - q_0) \tag{27}$$

in order to distinguish it from distributions computed in an approximate manner; $q_0$ is the sequence-ensemble average of the thermal average of $\langle \rho^1 | \rho^2 \rangle$. When the true distribution $P(q)$ has a single peak, we say that the system is replica symmetric [29] because $P(q)$ corresponds to an overlap matrix of the form shown in Figure 9, in which each replica is related symmetrically to every other replica.

We can identify the location $q_0$ of this single peak, for the homopolymer (homogeneous-sequence) limit $g^{(1)} = g_\lambda^{(1)}(T)$, $g^{(2)} = 0$, by computing the thermal average overlap of the 2-replica system with $g_t = g^{(1)}$ for each $t$. The distribution will be well-localized around the thermal average due to the clustering property [29] of the relevant single-replica "component" (the term component is discussed in Section V.B). This value of the overlap, $q_u$, is the totally uncorrelated value in the sense that the two replicas are not correlated by the presence of sequence fluctuations. Thus, for the homopolymer $g^{(1)} = g_\lambda^{(1)}(T)$, $g^{(2)} = 0$, we have $q_0 = q_u$. In the Appendix, we compute $q_u = \theta^2$.

We naively expect that for the random heteropolymer (random-sequence model) case, $g^{(1)} = g_\lambda^{(1)}(T)$, $g^{(2)} = g_\lambda^{(2)}(T)$, there will be some temperatures at which the peak in the true overlap distribution will shift towards values that are higher than $q_u$, i.e., $q_0 > q_u$. That is, we expect that increasing the scale of the fluctuation in the potential $g_t$ will cause the 1-d monomers to be found more often in positions where the potential is low than positions where the potential is high. This tends to increase the thermal average overlap, which gives the location $q_0$ of the peak of the true distribution $P(q)$.

In this work, we employ the replica method to test the stability of the realization of the overlap distribution that is localized at $q_u$. In Section III, we constructed the cuRSB overlap matrix (Figure 8), that predicts that the sequence-ensemble averaged overlap distribution should have the form

$$P_{cu}(q) = m\delta(q - q_u) + (1 - m)\delta(q - q_c). \tag{28}$$

where $q_c$ is the value of the thermal average overlap between two replicas that are constrained to have perfectly correlated motions of the 1-d monomers, i.e., $q_c = \theta$. The distribution $P_{cu}(q)$ is an approximation for the true distribution $P(q)$, which is useful because it allows for weight in a peak at $q > q_u$. We assume that if the variational procedure results in a weight at $q_u$ that is less than unity, i.e., $m < 1$, this suggests that the true distribution will exhibit a peak at some $q_0 > q_u$, indicating that the system is taking advantage of the sequence heterogeneity (quenched disorder). This approximate overlap distribution allows us to study melting and unzipping behavior in a way that takes some account of the enhanced overlap due to sequence heterogeneity.

**B. Homogeneous-sequence perfect-matching model treated with uRSB**

We apply cuRSB to the homopolymer (homogeneous-sequence) case, by setting $g^{(1)} = g_\lambda^{(1)}$, $g^{(2)} = 0$ in Eq. (A.10). The variational process selects the replica group-size $m = 1$, see Figures 10 or 11, for loop exponents $c = 2.115$ or $1.75$, respectively. Thus, two replicas of the homopolymer will show no tendency to move in unison and the overlap will have the uncorrelated value $q_u$, see Eq. (28). Fixing $m = 1$ results in a special case of cuRSB that we call uRSB; we still require that $\beta f_0$ is stationary with respect to $\theta$ and maximized (see Appendix A.3) with respect to $m$.

The uRSB "minimization" conditions, for $g^{(2)} = 0$, reduce to the single stationarity equation in Eq. (A.12), which is identical to the equation that would be obtained if we solved the homopolymer using a maximum-term method in which we select the value of the helix fraction that minimizes the system's free energy (see Appendix A.3). Thus, our results are consistent with methods not based on replicas.

In Figures 10 and 11, we also plot the values of the critical force $F_c$ and the helix fraction $\theta$ obtained from the uRSB solution. Also, we plot $F_c$ and the order-parameter at the critical temperature

$$T_c^{homo} = \frac{h}{s + R \ln\left\{\frac{1}{1 - A[\zeta(c) - 1]}\right\}} \cong T_{all}\left\{1 + \frac{R}{|s|} A[\zeta(c) - 1] + O(A^2)\right\}, \quad (29)$$

which can be derived from the stationarity equation, see Eq. (A.12), by setting $\hat{x} = 1$. The notation $\zeta(s)$ is the Riemann zeta function. The logarithmic term, which is positive, shows that loops provide a small entropic stabilization of the bound phase. The temperature $T_{all} = h/s$ is the location of the melting transition of the all-or-none version of our model. In the all-or-none model [10, 34], which is obtained by taking the $A \to 0$ limit of our model, the molecule may exist in only one of two microstates: (1) all bp's are in the helix state (2) all bp's are in the coil state. The location $T_{all}$ of the $1^{st}$-order melting transition of the all-or-none model is the temperature at which the helix propagation free energy equals the free energy of the free coil macrostate, i.e., $g^{(1)}(T) = 0$. We write $T_c^{homo}$ in terms of $T_{all}$ to emphasize that they are very close in numerical value; for the genome of λ-DNA $T_{all} = 90.5478$, $T_c^{homo} = 90.5505$ for $c = 2.115$, and $T_c^{homo} = 90.5481$ for $c = 1.75$ (significant figures chosen to discriminate amongst the temperatures).

To characterize the melting transition of the homopolymer, we say that the molecule is essentially a complete helix, $\theta = 1$, up to $T_{all}$. Here, loops emerge in the structure of the molecule, and the helix fraction falls from unity to

$$\theta(T_c^{\text{homo}}) = \frac{1 + \tilde{A}[\zeta(c) - 1]}{1 + \tilde{A}[\zeta(c-1) - 1]} \tag{30}$$

where $\tilde{A} = A e^{g^{(1)}/RT}$ and with the notation $\zeta(s)$ with $s \leq 1$, we mean positive infinity, see Appendix A.3 for derivation. So, if $c = 2.115$, $\theta(T_c^{\text{homo}}) = 0.9986 > 0$, i.e. the helix fraction drops discontinuously to zero at $T_c^{\text{homo}}$ and the melting transition is termed 1st-order. But, if $c = 1.75$, $\theta(T_c^{\text{homo}}) = 0$, i.e. the helix fraction drops continuously to zero at $T_c^{\text{homo}}$ and the melting transition is termed higher-order.

These transition orders are expected on the grounds of previous theory [9, 35], where it was demonstrated that the order of the melting transition of the homogeneous perfect-matching model of dsDNA depends upon physical estimates of the parameter $c$.

$$\begin{aligned} 2 < c : \quad & \text{1st-order phase transition} \\ 1 < c \leq 2 : \quad & \text{higher-order (continuous) phase transition} \\ c \leq 1 : \quad & \text{no true phase transition} \end{aligned} \tag{31}$$

If it were the case that $c \leq 1$, the helix fraction would decrease as a function of increasing temperature, but never fall to zero at any finite temperature.

Thus, our replica stationarity equations are consistent with the results of existing homogeneous PM theory. But if we look at Figures 10 and 11, we see only a small difference between the helix fraction curve for $c = 2.115$ versus $c = 1.75$. Due to the smallness of $A$, $\theta$ drops from unity to $\theta(T_c^{\text{homo}})$ over an interval of temperatures that is too small to see in these plots. Thus, on the temperature scale of these plots, and on the temperature scale of experiments, the homogeneous-sequence melting transition appears 1st-order due to the cooperativity provided by the parameter $A$. For this reason, it is likely that neither experimental melting curves, nor unzipping experiments, of long homogeneous DNA can determine the true value of the loop exponent $c$. The smallness of $A$ hides the value of $c$ from experimental determination.

### C. Random-sequence perfect-matching model treated with cuRSB

We find it interesting to apply the uRSB scheme to the random heteropolymer (random-sequence) case $g^{(1)} = g_\lambda^{(1)}(T)$, $g^{(2)} = g_\lambda^{(2)}(T)$. For both the Kafri and Fisher values of the loop exponent $c$, Figures 10 and 11 shows that the uRSB solutions cease above about $83°C$. In terms of the 2-replica picture, this means that below $83°C$ the true overlap $q_0 \cong q_u$, whereas above $83°C$, $q_0 > q_u$ and so uRSB is a poor approximation.

We now apply the cuRSB scheme to the random heteropolymer case. Specifically, we restrict the cuRSB equations to $m < 1$, calling the procedure cRSB. Here, we make $\beta f_0$ stationarity with respect to $\theta$ and $m$, then we check that the resulting stationary point is a local maximum in the $m$ direction. For $c = 2.115$ or $c = 1.75$, the results are shown in Figures 10 or 11. We see that cRSB succeeds where uRSB fails. Solution of the cRSB scheme indicated that cuRSB has selected $m < 1$, thus the approximate distribution $P_{cu}(q)$ in Eq. (28) has weight in the peak at $q = q_c$. We interpret this as meaning that for the true overlap distribution $P(q)$, $q_0 > q_u$, and the monomers of the 1-d polymer are localizing at sequence positions where $\delta g_t$ is low.

For both $c = 1.75$ and $c = 2.115$ (Figures 10 and 11), as $T$ is increased, the helix fraction of the random heteropolymer departs from unity at a lower temperature than for the homopolymer. In other words, the sequence fluctuations of the random sequence stabilize loops at temperatures where they are suppressed in the homogeneous sequence. Moreover, loops emerge in the random-sequence model well below $T_c$, whereas in the homogeneous model loops are suppressed until extremely close to $T_c$; these results are consistent with existing numerical results for random-sequence models of DNA [10]. For $c = 2.115$ $T_c = 95.33°C$, and for $c = 1.75$ $T_c = 95.14°C$; the computation of $T_c$ is explained in Figure 10. The model predicts that loops emerge at $83°C$, which is about $12°C$ below $T_c$. This is consistent with classic spectroscopic measurements on the λ genome, showing that the width of the melting transition is $10 - 15°C$ [2,3]; note that these measurements were obtained at a variety of salt conditions.

To compare our calculations directly with experiment, we measured circular dichroism (CD) melting curves of λ-phage genome in phosphate-buffered saline (PBS), see Figures 10 and 11. To obtain this data, the temperature was ramped at $5°C/\min$ over the interval $15 - 105°C$, we averaged over four runs, the concentration of the DNA was $100 \mu g/mL$, and the device is a Jasco J-710 Spectropolarimeter with a PTC-378 W Jasco temperature controller. We compared two molecules, (1) the molecular construction used in unzipping studies, that contained a hair-pinned λ genome [4,36] and (2) the naked λ genome (New England Biolabs). We found that the temperature at which the reading decreased from the low-T value (emergence of loops) and the temperature at which the reading met the high-T value (separation of strands) were very similar between the two molecules (data not shown). In Figures 10 and 11, we show only the data for the naked λ genome. The agreement between theory and experiment is extremely good. Not only does our theory predict the temperature at which loops emerge, but it predicts the melting temperature $T_c$ to within 0.8% on the Kelvin scale. The predicted $T_c$ is $95°C$ and the observed is approximately $98°C$, see Figures 10 and 11. This agreement constitutes quantitative accuracy.

Regarding the self-consistency of the infinite-chain approximation, Poland and Scheraga [37] state that, at zero force, the infinite-chain approximation is valid when most of the denaturation occurs through the formation of internal loops rather the lengthening of the unbound sequence; this occurs when $N/L_{helix} >\sim 100$ (meaning greater than or of the order of). In our calculation, near $T_c$, $L_{helix} < 100$ (see Figure 10), so for the λ-genome $N/L_{helix}$ is at least 480. Thus, the infinite-chain approximation is valid.

In the next section, we will focus on the following feature of the results: the presence of sequence heterogeneity smoothes the fall of the helix-fraction from unity to zero. This smoothing effect is so strong, that for the parameter $c = 2.115$, the random heteropolymer has a higher-order melting transition whereas the homopolymer has a true 1$^{st}$-order melting transition. For $c = 1.75$, the homopolymer predicts a true higher-order melting transition that appears 1$^{st}$-order because $T_{all}$ is extremely close to $T_c^{homo}$, and this apparent 1$^{st}$-order transition is converted into a high-order transition by the sequence heterogeneity. For both values of $c$, the presence of sequence heterogeneity raises the apparent order of the melting transitions, and this effect shows up in an effective value of $c$, that controls the behavior of the helix fraction and the critical force in the neighborhood of $T_c$. Because of the availability of the new experimental techniques of single-molecule forced unzipping, we focus on the critical force in the next section.

## V. Discussion

### A. The effective loop exponent and the order of the melting transition

#### 1. Critical force scaling near $T_c^{homo}$ for the homogeneous-sequence model

As temperature is increased, the homopolymer remains a complete helix up to approximately $T_{all} = T_c^{homo} - O(A)$. Given that the resolution of Figures 10 and 11 is about $1°C$, the shape of the critical force curve near $T_c^{homo}$ is determined by the all-or-none model. If we set $g^{(1)}(T) = g_F(T, F_c)$, and expand about $T_{all}$, we obtain

$$F_c = \frac{k_B T}{b}\left[ e^{-\frac{b}{a}\beta g^{(1)}(T)} - 1 \right] = \frac{k_B T}{(ab)^{1/2}} \left(\frac{|s|}{R}\right)^{1/2} (T_{all} - T)^{1/2}, \tag{32}$$

which gives the shape of the critical force curve in the neighborhood of $T_c^{homo}$. From Eq. (32), we see that the critical force scales like

$$F_c : |T - T_c|^\alpha \tag{33}$$

with $\alpha = 1/2$ exactly, to the extent that $T_{all} \cong T_c^{homo}$ and $T_c = T_c^{homo}$.

But, if we were to plot $F_c$ for $T_{all} < T < T_c^{homo}$, theory predicts that the shape of $F_c$ is determined by the free energetics of the loops. Specifically, in a homogeneous-sequence PM (perfect-matching) model, in which volume exclusion effects in the loop sequences are incorporated through the value of $c$ [20], Mukamel and Shakhnovich showed that above $T_{all}$ and very near $T_c^{homo}$, the scaling law Eq. (33) holds but with $\alpha = 1/2\eta$, where $\eta = \min(1, c-1)$, and $T_c = T_c^{homo}$. Notice that if $c \geq 2$, then both the loop-dominated scaling law, with $\alpha = 1/2\eta$, and the loop-less scaling law, with $\alpha = 1/2$ [see Eq. (32)], predict that the critical force grows with the square root of the distance below $T_c^{homo}$, at least to the extent $T_{all} \cong T_c^{homo}$. But if $c < 2$, then the critical force curve will exhibit a crossover as $T$ is increased passed $T_{all}$, because the exponent of the loop-full scaling law falls below the exponent of the loop-less law [Eq. (32)]. Thus, even if one could find a sequence of dsDNA that is well-approximated by the homogeneous PM model, it is unlikely that an experiment would be able to resolve this crossover because $T_{all} \cong T_c^{homo}$. Thus, it is unlikely that experiment, either spectroscopic melting curves or forced unzipping, can discriminate between different values of the loop exponent.

## 2. Critical force scaling near $T_c$ for the random-sequence model

Similar to Poland and Scheraga theory [9], we encode a single parameter with the manner in which sequence heterogeneity raises the order of the melting transition. We take the critical force curve near $T_c$ of the random-sequence model, and fit it to the homogeneous-sequence model, i.e. $\alpha = 1/2\eta$, and extract the value of the loop exponent $c^{fit}$ that solves $\eta = \min(1, c^{fit} - 1)$. For the random PM model at $c = 2.115$, the fitted value of the loop exponent is $c^{fit} = 1.57$; because $c^{fit} < 2$, the melting transition is higher order, according to Poland-Scheraga theory [9], see Eq. (31). For the random PM model at $c = 1.75$, $c^{fit} = 1.56$. Thus, for both values of $c$, $c^{fit} < c$ quantitatively encodes the result that sequence heterogeneity raises the order of the melting transition.

## 3. The order of the melting transition for the random-sequence model

We corroborate the above interpretation of $c^{fit}$ with an analytical result for the distribution of loop lengths. For the infinite-chain, homogeneous, PM model, of all the loops that are present in the chain, the fraction of loops $n_\ell$ that have length $\ell$ is a fluctuating variable with mean value $\langle n_\ell \rangle = x^\ell \ell^{-c} / [\phi(x,c) - 1]$, where $\ell \geq 2$, $x$ is a fugacity that sets the helix fraction, and $\phi(x,c)$ is the polylogarithm function [see Eq. (A.8) ]. For the infinite-chain, random-sequence, PM model, we can obtain $\langle n_1 \rangle$, averaged over the sequence-ensemble, resulting in

$$\overline{\langle n_1 \rangle} = \frac{x^1 1^{-mc}}{\phi(x,mc)-1} \tag{34}$$

which is the distribution of loop lengths of a typical sequence in the sequence ensemble. Eq. (34) can be obtained from the replica free energy [Eq. (23)] by differentiation with respect to a force that tends to increase the number of loops of length $\ell$. We call $c^{ef} = mc$ the effective loop exponent, as it governs the distribution of loop lengths for a typical random sequence. For example, we checked that the average loop length obtained from distribution $\overline{\langle n_1 \rangle}$ agrees with $L_{loop}$ obtained in Section III.C. From Figures 10 and 11, we can read off that at $T_c$, $m = 0.629$ for either $c = 2.115$ or $1.75$, and thus $c^{ef} = 1.330$ or $1.10$, respectively. These values are in broad agreement with the values $c^{fit} = 1.57$ or $1.56$, respectively, where the quantitative discrepancy may be due to the fact that the fitting interval is $4\,°C$ wide, whereas we have only given $c^{ef}$ at $T_c$.

Since this $c^{ef} = mc$ appears in our equations for the helix fraction and the free energy (per bp) [see Eq. (A.11) and Eq. (23)], we conclude that $c^{ef}$ determines the order of the melting transition as higher order (not 1st order), and determines the shape of the critical force curve $F_c$ immediately near $T_c$. From homogeneous PM theory, if $1 < c \le 2$, the higher-order transition [see Eq. (31)] may be classified more specifically in terms of the exponent of the scaling relation between the helix fraction and temperature near $T_c$

$$\theta : (T_c - T)^\beta \tag{35}$$

where $\beta = (2-c)/(c-1)$ as derived by M. E. Fisher [10]. In our random-sequence theory, the helix fraction $\theta$ is the sequence-ensemble average of the thermal average of the fluctuating helix fraction, and can be obtained as

$$\theta = \overline{\langle \theta[\rho]\rangle_g} = \frac{\partial}{\partial h}\overline{f_g} = \frac{\partial}{\partial h}\overline{-\frac{RT}{N}\ln Z_g} \qquad (36)$$

as can be derived from the partition function of the 1-dimensional polymer [see Eq. (10)], and checked against the free energy per bp in Eq. (23); $h$ is the enthalpic component of $g^{(1)}(T)$, $\theta[\rho]$ is the helix fraction for microstate $\rho$, and $\langle ...\rangle_g$ is the equilibrium thermal average for a fixed sequence $\{g_t\}$. Thus we can say that the transition is continuous (higher order), but we can also give the order-parameter's scaling exponent $\beta = 2.04$ or $8.90$, which correspond to $c^{ef} = 1.33$ or $1.10$, which correspond to $c = 2.115$ or $1.75$, respectively. Consequently, for both for $c = 2.115$ and $c = 1.75$, the $2^{nd}$ derivative of the free energy density $\partial\theta/\partial T = \partial^2/\partial T\partial h$ is continuous at the melting transition. In the extended Ehrenfest classification [6], this phase transition would be classified as $3^{rd}$-or-greater order. We may generalize by giving the lowest-order derivative at which a divergence appears for arbitrary $c^{ef}$

$$\begin{array}{ll} 2 < c^{ef} & \text{1st derivative has jump discontinuity} \\ 3/2 < c^{ef} \leq 2 & \text{2nd derivative has divergent discontinuity} \\ c^{ef} = 3/2 & \text{2nd derivative has jump discontinuity} \\ c^{ef} < 3/2 & \text{3rd-or-greater derivative has discontinuity} \end{array} \qquad (37)$$

Importantly, homogeneous-sequence theory makes the prediction that for $c = 2.115$, the melting transition is $1^{st}$-order, having a jump discontinuity in the helix fraction. For $c = 1.75$, homogeneous PM theory predicts that $\beta \approx 1/3$, so that $\partial\theta/\partial T$ diverges at $T_c$. Our random-sequence PM theory contradicts both these statements, predicting that both $\theta$ and $\partial\theta/\partial T$ are continuous across the transition. Thus, our random-sequence theory is consistent with experimental melting curves of phage λ, see [2] [3] and Figures 10 and 11, while the homogeneous-sequence theory is not.

**B. Previous theoretical efforts on the random-sequence perfect-matching model**

We acknowledge that other calculations have been devised to study the zero-force melting of random-sequence models of long natural dsDNA. Poland and Scheraga [37] treated the infinite, random-sequence, perfect-matching (PM) model with an annealing-sequence approximation developed by Lifson and Allegra [13, 14], in which the helix-propagation free energies $g_t$ are considered as mechanical variables that anneal hand-in-hand with the conformational microstate of the molecule [10]. While such methods are analytical, these methods are less favorable because they do not, in general, result in the thermodynamics of a typical random sequence. Additionally, the annealing-sequence methods have been shown to give poor approximations to the exact results (Lehman and McTague [12]) in the case of the nearest-neighbor Ising model [10, 15]. Our calculation is advantageous, because the replica method [29] offers an analytical route to perform the sequence-ensemble averaging operation exactly, at least in principle. In practice, approximation schemes involved in the replica method reduce the exactness of the results, but we argue that (a) the results are "more exact" than annealing-sequence methods (b) the replica method offers the possibility of reducing the severity of the approximation. Regarding specific results of annealing-sequence methods, Poland and Scheraga [37] found the dubious result that, for $c = 1.75$, the typical length of a helix sequence was the same for the homogeneous-sequence and random-sequence models.

Contrariwise, we find that near $T_c^{homo}$, the helix length of the random-sequence model is orders of magnitude shorter, see Figure 11. But the most significant advantage of our calculation, is that our theory outputs an effective loop exponent $c^{ef}$, that analytically controls the shape of $F_c$ near $T_c$, and thus cleanly classifies the melting transition as a type of phase transition.

In another theoretical effort, Poland and Scheraga have numerically studied the finite-chain, random-sequence, PM model. They found transition widths consistent with our results [10, 37]. Since this model is finite, it does not yield a classification of the phase transition that results in the infinite-chain limit. As an alternative to an infinite-chain calculation, one may study how the properties of the finite model change as the system size is increased, but to our knowledge, this has not been done.

## C. Experiment analyzed with theory

### 1. Unzipping the 1st 1500 bp's and the all-or-none model

In a multiplexed single-molecule experiment to determine the temperature-force phase diagram for unzipping the first 1500 bp's of the 48502 bp genome of phage λ, Danilowicz et al. presented an all-or-none model (finite-chain) for the critical force curve [4]. The free energy per bp of the all-helix microstate was $g_{fit}^{(1)}(T) = h_{fit} - Ts_{fit}$, with $s_{fit} = -20.6$ $cal/mol$ obtained from averaging over the first 1500 bp's at the left end of the l-strand, and $h_{fit} = T_{1/2}s_{fit} = -7.5$ $kcal/mol$, where $T_{1/2} = 90.9$ °C was determined as the mid-transition point of the CD (circular dichroism) melting curve of the entire genome of phage λ. This value of $T_{1/2}$ was assumed a good approximation for the melting temperature of a molecule composed of the first 1500 bp's, because the width of the transition of the entire genome is $\approx 15\ K$ while $T_{1/2} \approx 364\ K$, so the error is about 4%.

In Figure 13, we show the critical force curve that results from these parameters (plus signs); we have ignored the elastic term of the model for the free energy of the unbound section used in that work [4]. At each temperature $T$, the critical force $F_c$ is computed by solving $g_F(T, F_c) = g_{fit}^{(1)}(T)$, see Eq. (6). The data point at $F_c = 0$ is the melting temperature $h_{fit}/s_{fit}$, which has value $T_{1/2}$. At each point on the $F_c(T)$ curve, the model is evenly distributed between the all-helix and all-coil state, so the helix fraction is 0.5. Note that the equilibrium critical force $F_c$ predicted by the all-or-none model is identical to the prediction of the uncorrelated random zipper model, if the free energy (per bp) of the all-helix state of the two models is the same. Also, the prediction for $F_c$ of the all-or-none model is identical to the prediction of our infinite-chain RHP perfect-matching model for $20 \leq T \leq 50\ °C$, but we do not discuss the latter model here because the infinite-chain limit is not appropriate for experiments done on $1500\ bp$ sequences.

We also show the results of our own calculation of the critical force curve predicted by the all-or-none model, see Figure 13. For the free energy per bp of the all-helix microstate, we used $g_\lambda^{(1)}(t, t'; T) = h_\lambda(t, t') - T s_\lambda(t, t')$, which is computed with Eq. (13) in Section II.B, but we restricted the average to the portion of the genome between sequence positions $t$ and $t'$. As above, the critical force is obtained from $g_F(T, F_c) = g_\lambda^{(1)}(t, t'; T)$, and the melting temperature is computed as $T_{all}(t, t') = h_\lambda(t, t')/s_\lambda(t, t')$. For the sequence $[1, 1500]$, our computed critical force is roughly 1.5 pN higher than the critical force computed by Danilowicz et al. [4]. In our calculation, $h_\lambda(1,1500) = -9.46 \; kcal/mol$ and $s_\lambda(1,1500) = -26.0 \; kcal/mol$, whereas Danilowicz et al. obtained different values because they used a different nearest-neighbor parameter set and neglected the corrections for sodium concentration. This discrepancy is important, because the Danilowicz calculation suggests that the experimentally reported critical force $F_c^{report}$ is a good estimate of the true equilibrium critical force of the system $F_c$, which we may define as the force at which thermal equilibrium will distribute an ensemble of experimental molecular constructions evenly between the all-helix and the all-coil state. Contrariwise, our calculation – which predicts $F_c$ via the all-or-none model – suggest that on $24 \leq T \leq 35 \; °C$, the reported force $F_c^{report}$ is about $2 \; pN$ less than $F_c$. This discrepancy may reflect that the experiment is complicated by kinetic barriers that prevent the attainment of equilibrium on the experimental time-scale. These barriers could be, e.g., of the type embodied in the random zipper model[32]. Alternatively, the discrepancy between $F_c^{report}$ and the all-or-none model's prediction for $F_c$ may reflect interactions that are important in the real system, that are not included in the Blake-Delcourt nearest-neighbor parameters [22], which are obtained from experiments at much higher temperatures.

An additional failure of the all-or-none model parameterized with the Blake-Delcourt nearest-neighbor model, is the inability to explain the drops in the experimentally reported critical force $F_c^{report}$ that occurs near $25 \; °C$ or $40 \; °C$. We speculate that these discrepancies between model and experiment may be due, as above, to (a) kinetic barriers preventing experimental sampling of equilibrium (b) interactions not accounted for in the high-temperature nearest-neighbor parameters of Blake and Delcourt [22].

Despite these discrepancies, our results of the all-or-none model follow expected trends in stability, see Figure 13. At a given temperature $T$, the sub-sequence labeled "weak middle" has the least stable helix-state free energy, the AT-rich subsequence is more stable, and the GC-rich subsequence is even more stable. At a given temperature $T$, the critical force $F_c$ to unzip the model parameterized on these subsequences follows this same trend, and this serves as a consistency check for our calculation. The values of the melting temperatures $T_{all}$ also follows this trend.

*2. Unzipping the entire 48502 bp genome and perfect-matching models*

Because the finite-chain all-or-none model reproduces the experimental unzipping of the first 1500 bp's on $24 \leq T \leq 35\,°C$, the all-or-none model is the null model for the critical force curve of the entire genome. But the all-or-none model, trained on the entire genome, will fail to model the entire genome because the width of the melting transition of the model was computed (by us) to be $\square\, 0.01\,°C$ (data not shown), much smaller than the width of the experimental CD melting curve. The all-or-none model is a two-state model, so the helix fraction at zero force is $\theta_{all}(T) = \left[1 + u_h^{-N}\right]^{-1}$, where $u_h = exp(-g/RT)$, and $g$ is the free energy per bp of the all-helix microstate. It is the dependence on $N$ that makes the melting transition very narrow. Thus, both the finite-chain and the infinite-chain all-or-none models are poor models for the unzipping phase diagram of the entire genome.

We may also rule out the infinite-chain zipper model as a model for the entire unzipping phase diagram of the genome of phage-λ. It was shown in [32] that both the homogeneous-sequence and random-sequence, infinite-chain, zipper models have the same critical force curve $F_c(T)$ as the all-or-none model, for which $F_c = 0$ at the melting temperature $T_c = 90.55 °C$, see Figure 13. This prediction disagrees with experimental CD melting curves, as shown in Figures 10 and 11. We cannot rule out a finite-chain zipper model, but these results suggest that loops are necessary to model the entire λ-phage genome.

To include loops, one may propose a homogeneous-sequence perfect-matching (PM) model. In Section IV.B, we showed that the width of the melting transition of the infinite-chain homogeneous PM model is too small to model the experiment. Similarly, the finite-chain homogeneous PM has been shown, via matrix calculations, to have a narrow melting transition; for chain-length $N = 5000$ bp's and cooperativity parameter $A = 10^{-3}$, the width was about $1 °C$ [9]. Since λ-DNA has higher $N$ and lower $A$, the melting transition of the homogeneous PM model for λ-DNA, will have an even smaller width. Thus, both loops and sequence heterogeneity are necessary to model the width of the melting transition – and thus to model the unzipping phase diagram – of the entire λ-phage genome.

We now compare the random PM model to a recent set of unzipping experiments, in which the extension of an individual copy of the earlier described molecular construction [36] is recorded as the applied force is first increased, and second decreased, at a constant rate (data not published). The applied force is held constant for $2 \sec$, increased by $0.5 pN$, etc., until the maximal extension is observed, and then the force trajectory is reversed. The extension at the maximal value of the force confirms the unzipping of the entire 48502 bp genome, and the extension at the minimal force suggests the complete rezipping.  As a function of temperature, the force at which unzipping occurs, and the force at which rezipping occurs, is plotted in Figure 14.  As expected from the landscape picture of unzipping present in the uncorrelated random zipper model, the unzipping and rezipping forces bracket the equilibrium critical force $F_c$ predicted by the perfect-matching RHP model.  Note that at the temperatures of the experiment, the perfect-matching RHP reduces to the all-or-none model, because no loops are present.  It is difficult to comment further on the relationship between our model and this experiment, because the forces reported in the experiment will depend upon the sequence dependent kinetic barriers to unzipping and rezipping, and the rate at which the applied force is changed.  Due to the barriers and non-zero rate of change of the applied force, the extension of the molecule is probably not governed by thermal equilibrium.

## VI. Conclusion

We have calculated the phase diagram of the random-sequence, infinite-chain, perfect-matching (PM) model for the single-molecule forced unzipping of the 48502 bp genome of phage lambda.  To our knowledge, this is the first calculation of the unzipping phase diagram of a dsDNA model with loops and sequence heterogeneity.  This calculation was performed analytically with the replica method.  To our knowledge, this is the first time that replica method has been applied to a model for dsDNA.  In the phase diagram, the critical force curve defines a line of 1$^{st}$-order phase transitions.  The line ends at zero-force where there is a higher-order melting transition.  The critical force curve near $T_c$, has an experimentally observable shape that gives evidence that the melting transition is higher-order (continuous). Specifically, our theory predicts that the transition is a 3$^{rd}$-or-greater order phase transition.

We contrast the random-sequence critical force curve against the results for the homogeneous-sequence model. In the homogeneous model, loops are suppressed until less that $0.01°C$ below $T_c$. Consequently, for experimentally observable temperatures near $T_c$, the homogeneous-sequence Fc has a shape that is determined by the all-or-none model, giving a scaling law $F_c : |T - T_c|^\alpha$ with exponent $\alpha = 1/2$ almost exactly. Thus, the true 1$^{st}$-order phase transition associated with the Kafri et al. value of the loop exponent ($c = 2.115$) might be present but it is not experimentally observable. Contrariwise, the random-sequence model shows loops $12°C$ below $T_c$. The presence of loops cause the critical force curve to deviate substantially from that of the homogeneous-sequence model. The resulting $T_c$ is $\approx 5°C$ higher, and the critical force scaling law has exponent $\alpha = 0.881$ or $0.899$ for $c = 2.115$ or $1.75$, respectively, indicating a slightly sub-linear temperature dependence. Motivated by the presence of loops, we fit the random-sequence critical force curve to the scaling law $F_c : |T - T_c|^\alpha$, with $\alpha = 1/2\eta$ and $\eta = \min(1, c^{fit} - 1)$, obtaining a fitted value $c^{fit}$ of the loop exponent. Since $c^{fit} = 1.57$ or $1.56$ for $c = 2.115$ or $1.75$, respectively, Poland-Scheraga theory states that the melting transition is higher-order. These exponents are experimentally accessible observables that can be determined for the 48502 bp λ–phage genome. The experimental validation would support our claim that $c^{ef}$ predicts that the melting transition is 3$^{rd}$-or-greater order.

The experimental measurement of $F_c$ near $T_c$ is hindered by the denaturation of proteins involved in the molecular attachment of the genome to the plastic bead that responds to the magnetic tweezers. Due to this protein, the unzipping studies conducted in phosphate-buffered saline (PBS) are limited to temperatures below about $55°C$. But the melting temperature of dsDNA depends strongly on salt conditions. For example, the genome melts at about $75°C$ in a TRIS solution (data not shown). Thus, it is likely possible to find a solution with ionic conditions that lower the melting transition of the genome to temperatures where the molecular construction is stable.

To our knowledge, this is the first work in which the replica method has generated results with semi-quantitative agreement with experiment. Our calculation contains no adjustable parameters, and our predicted melting temperature agrees with the experimentally determined value to $0.8\%$ in the Kelvin scale.

We now compare our results to those for the Peyrard-Bishop (PB) model, in which loops are modeled as gaussian chains. The original PB model [38] corresponds to the PM model with $c = 3/2$; the homogeneous-sequence PB model has a higher-order melting transition, consistent with Poland-Scheraga theory. Recently, Cule and Hwa studied the variable-stiffness PB model [39], in which the gaussian-chain stiffness coefficient depends the loop configuration. The variable stiffness feature does not make the transition 1st-order, but only sharpens the transition [40]; it appears that the variable-stiffness parameter is analogous to the cooperativity parameter $A$ of the PM model. To the variable-stiffness PB model, Cule and Hwa added sequence randomness, finding that both variable-stiffness and sequence randomness where necessary to generate multi-step melting curves for sequences of approximately 3000 bp's [40]. Despite the multi-step behavior found for intermediate-length sequences (3000bp's), the authors speculated that the melting curve would become smooth as the length of the random sequence becomes large; this speculation is consistent with our results. Thus, the speculated behavior of the random-sequence, variable-stiffness, PB model is consistent with our results for infinite sequences: there is a phase transition of the higher-order (continuous) type.

## Acknowledgements

C.B.R. acknowledges support from the Ford Foundation Pre-doctoral Fellowship, the Harvard Graduate Prize Fellowship, and the Harvard Merit Fellowship. We thank Maxim Frank-Kamenetskii and David Nelson for helpful comments on this work. We thank Dima Lukatsky for comments on the manuscript.

## A. Appendix

### 1. No degenerate ergodicity-breaking

*(i) Introduction*

In the present work, we employ the replica method in order to compute the sequence-ensemble average of the free energy. The method requires an ansatz for the overlap matrix order parameter, e.g. see Figure 8. Below, we make remarks in order to interpret our particular choice for this ansatz. In particular, we demonstrate that ergodicity-breaking is not to be expected on regions of the phase diagram where the replica method is capable of detecting it.

*(ii) Heuristic argument*

Consider the 1-d polymer at fixed finite $N$, some $T$, and $F = 0$, with some particular sequence which is randomly generated at $g^{(1)} = g_\lambda^{(1)}(T)$, $g^{(2)} = g_\lambda^{(2)}(T)$. We define a dynamics for the system as a Monte-Carlo (MC) evolution. In this dynamics, we allow two types of "moves". In the first type, we attempt to move a randomly chosen monomer to an adjacent position. In the second type, we chose a position at random. If there is a monomer there, we attempt to remove it. If there is not a monomer there, we attempt to insert a monomer. Moves are accepted with a probability that is the ratio of Boltzmann weights for the system before and after the move. We assume that these microstate transition rules respect broken ergodicity if it exists.

We now define ergodicity-breaking with respect to this dynamics, using the language of Palmer [41]. We call the microstate space of the model $\Omega$ and define a component as a subset $\Omega^I$ of $\Omega$ such that the time-scale for escape from region $\Omega^I$ is much longer than the time-scale on which the model reaches equilibrium within $\Omega^I$. We say that the model has components with absolute confinement if at each $N$ we can identify a subset of components such that as $N$ grows, the escape time-scale for members of this subset increases unboundedly while the time-scale for reaching equilibrium within components in this set is bounded as $N$ grows. We say that the model breaks ergodicity at parameters $(T, F)$ if there is at least one absolutely-confined component $\Omega^I$ that is a proper subset of $\Omega$. Additionally, we say that the model exhibits thermodynamic ergodicity-breaking if the absolutely-confined component makes a thermodynamically important contribution to the partition function. Finally, we say that the model exhibits degenerate thermodynamic ergodicity-breaking if there are multiple thermodynamically important and absolutely-confined components. We imagine boundary regions between components as constituting large free energy barriers. In what follows, we will only be concerned with thermodynamically-important and absolutely confined components, which we will simply call components. We will only be concerned with ergodicity-breaking of the thermodynamic type, which we will simply call ergodicity-breaking. We focus our discussion on the number of components; if there are multiple components, the model has degenerate ergodicity-breaking.

We now present expectations about the presence or absence of degenerate ergodicity-breaking at different temperatures.

At high $T$, we expect each strand of the double-stranded model to have free coil statistics. We base this expectation on the existence of a finite temperature phase transition for a thermally melted homogeneous-sequence model with loop exponent $c > 1$ [9]. Specifically, if we model a long random sequence of DNA with extremely high GC-content as a homogeneous-sequence perfect-matching model, theory predicts a phase transition at some finite $T_{GC}$. Consequently, our model for the sequence of $\lambda$-DNA must melt at some $T < T_{GC}$. Thus, at high $T$, both the free energy per bp and the bound-state fraction should zero, up to finite $N$ fluctuations. If we call a particular arrangement of 1-d monomers a loop structure, then a single loop structure will dominate. Thus we expect there to be a single component, and thus there is no degenerate ergodicity-breaking.

As we lower $T$, we expect there to be a value, call it $T_c$ such that for $T < T_c$, the Gibbs equilibrium state [41] is dominated by one or many components that have a free energy density (i.e. per bp) that is less than zero. Again, this is suggested by the behavior of homogeneous-sequence models; at low enough $T$ our model should exist as a single helix, here the homogeneous-sequence model is accurate, and will have a negative free energy density, see Figure 5. In each component $\Omega^I$ we can say the following. The free energy density is negative (and of similar magnitude) and so the distribution of loop lengths has exponential decay for large loops, see Eq. (5). Thus, there is some cutoff, $\ell_{cut}$, above which there are essentially no loops. Also, the distribution of the length of the unzipped section has exponential decay. Thus the helix fraction must be non-zero. So we can identify $T_c$ as the largest temperature below which the helix fraction of the Gibbs state is non-zero. The value $T_c$ is the temperature above which the strands are separated, except perhaps for a thermodynamically unimportant number of helix-state bp's. We can call $T_c$ a critical temperature, in the sense of a divergent correlation length (typical loop size), if the melting transition is higher order (continuous).

We would like to know the number of these components $\{\Omega^I\}$ at low T. We hypothesize that there are many, i.e. that the model exhibits degenerate ergodicity-breaking, and consider the consequences.

We divide the 1-d polymer into subsystems by partitioning sequence-position space into intervals of size $l_{cut}$ and labeling them $b \in \{1,...,M\}$. The microstate of sub-system $b$, $\gamma_b$, is a particular arrangement of monomers on interval $b$. We remove the interaction between subsystems and consider the nature of $\Omega$ under a hamiltonian of the form $\sum_b h_b^{ef}(\gamma_b)$. Under the hypothesis that the whole system exhibits degenerate ergodicity-breaking, each independent subsystem should also exhibit degenerate ergodicity-breaking; we assume that free or bound boundary-conditions of the 1-d polymer does not affect the number of components. In the microstate space of subsystem $b$, call it $\omega_b$, there are multiple components $\omega_b^I$, each labeled by index $I$ and having an escape time-scale that grows with $\ell_{cut}$. The thermodynamics of the collection of subsystems is dominated by $\bigcup_{\{I_b\}} \otimes \prod_b \omega_b^{I_b} = \Omega_{thermo} \subset \Omega$.

We now consider a restricted version of model $\sum_b h_b^{ef}(\gamma_b)$ in which the microstate of the whole system, $\Gamma$, is confined to a sub-region $\otimes \prod_b (\omega_b^{I_b} \cup \omega_b^{J_b})$ such that each subsystem is restricted to the union of two particular components, $\omega_b^{I_b}$ and $\omega_b^{J_b}$, that are connected by the dynamics. We map this system to an Ising model by, for each subsystem $b$, assigning a spin state to each of the two components $I_b$ and $J_b$; we consider all such assignments. For each subsystem $b$, the difference in the constrained free energy on $\omega_b^{I_b}$ and $\omega_b^{J_b}$, under $\sum_b h_b^{ef}(\gamma_b)$, maps to a random magnetic field in the Ising description. We now add ferromagnetic coupling $J$ between neighboring spins with magnitude of order $\log \ell_{cut}$, obtaining a one-dimensional random field Ising model (RFIM). By adding the ferromagnetic interaction, we have constructed a model (d=1 RFIM) that is more cooperative than the original model $H_g$ restricted to $\otimes \prod_b (\omega_b^{I_b} \cup \omega_b^{J_b})$. But, the d=1 homogeneous Ising model is more cooperative than the d=1 RFIM. Since the Landau argument [16] tells us that the d=1 homogeneous Ising model does not have an ordering transition at $T > 0$, neither does the RFIM, and thus neither does $H_g$ restricted to $\otimes \prod_b (\omega_b^{I_b} \cup \omega_b^{J_b})$.

Because $H_g$ restricted to $\otimes \prod_b (\omega_b^{I_b} \cup \omega_b^{J_b})$ does not have an ordering transition at finite $T$, the behavior of the high-T macrostate persists to all $T > 0$: each subsystem $b$ acts essentially independently of its neighbors, inter-converting between $\omega_b^{I_b}$ and $\omega_b^{J_b}$ with finite time-scales. Thus, there is no whole-system microstate $\otimes \prod_b \omega_b^{I_b}$ with escape time growing with $N$. So, $H_g$ on $\otimes \prod_b (\omega_b^{I_b} \cup \omega_b^{J_b})$ does not exhibit degenerate ergodicity-breaking at non-zero $T$.

We now consider another restricted version of model $\sum_b h_b^{ef}(\gamma_b)$ in which $\Gamma$ is confined to the sub-region $\otimes \prod_b (\omega_b^{I_b} \cup \omega_b^{J_b}) \cup \omega_b^{K_b}$, where $\omega_b^{K_b}$ is dynamically connected to $\omega_b^{I_b}$ or $\omega_b^{J_b}$. Again, this restricted model is equivalent to a d=1 RFIM after we have assigned, for each subsystem $b$, two of the three components to one spin state, and the third component to the second spin state. Similar arguments as above show that $H_g$ does not exhibit degenerate ergodicity-breaking when restricted to

$$\otimes \prod_b (\omega_b^{I_b} \cup \omega_b^{J_b}) \cup \omega_b^{K_b}.$$

We continue this procedure until we arrive at the statement that $H_g$ does not exhibit degenerate ergodicity-breaking when restricted to $\otimes \prod_b (\bigcup_{I_b} \omega_b^{I_b})$, which is $\Omega_{thermo}$. This contradicts the hypothesis of many components in $\Omega$ under $H_\varepsilon$. We conclude that for $T < T_c$, $\Omega$ contains only one component.

Thus, there can be no degenerate ergodicity-breaking for $T < T_c$.

*(iii) The overlap distribution has a single peak*

In replica theory [29], if a model does not exhibit degenerate ergodicity-breaking, then the corresponding distribution of the 2-replica overlap exhibits a single peak. The converse is also true. This relationship can be understood with the following argument.

For a typical sequence in the sequence-ensemble, if one replica with this sequence has a single component, then two non-interacting replicas of this sequence form a combined system that has a single component, see Eq. (26). Consequently, the 2-replica overlap $\langle \rho^1 | \rho^2 \rangle$, which is an intensive property of the two replica system, has contributions from $O(N)$ independent random numbers (i.e., has the clustering property [29]), so that fluctuations are limited to the scale $1/N^{1/2}$ (see Note 1 of Chapter 1 of [42]).

## 2. Derivation of $Z_{group}$

### (i) Explicit expression for $Z_{group}$

We first obtain a formal expression for the entropy term of the cuRSB free energy [see Eq. (23)]. At large integer $n$ and $m \in \{1, \ldots, n\}$, we apply the cuRSB constraints to the n-replica system $H_n$ and turn off the term that couples the replicas by setting $g^{(2)} = 0$. We obtain a system of $n/m$ independent groups, each group composed of $m$ perfectly correlated chains. We obtain the macrostate of this system as a function of $\theta$ and $m$ by computing the right-hand-side of

$$e^{S_n[Q^c]} \cong [Z_{group}(N, p, m, g_F)]^{n/m} \tag{A.1}$$

where $p = \lceil N\theta \rceil$, where $\lceil \cdots \rceil$ is the ceiling function, and the approximate equality becomes exact in the thermodynamic limit. For $\theta > 0$, we have

$$Z_{group} = \sum_{1_1=1}^{\infty} \cdots \sum_{1_p=1}^{\infty} \sum_{1_{p+1}=0}^{\infty} \delta(N, \sum_{i=1}^{p+1} 1_i) u_F(1_{p+1}) \prod_{i=1}^{p} u(1_i)^m \tag{A.2}$$

where $\delta(a, b)$ is the Kronecker delta, and

$$u(1) = \begin{cases} 1 & , \ l=1 \\ \dfrac{A}{c} e^{\dfrac{g^{(1)}}{RT}} & , \ l \geq 2 \end{cases} \quad (A.3)$$

is the partition sum of a bound sequence constrained to length $\ell$, where the free energy of the reference state has been shifted by $g^{(1)}$ units so that the stacking energy and main-chain entropy of the helix propagation free energy can be put in the energetic term of $E_n[\langle \rho | \rho \rangle]$ [see Eq. (19)]. The partition function of an unbound sequence of length $\ell$ is

$$u_F(1) = e^{-\dfrac{g_F}{RT} 1}. \quad (A.4)$$

For the computation of the fixed $N$ partition function $Z_{group}$, we choose the Leontovich method, applied, e.g., by Lifshitz to the computation of the entropy of a gaussian chain constrained to a particular realization of its density field [25]. In the first step, we allow $N$ to fluctuate against a force represented by fugacity $x$. Thus, we construct the grand-canonical partition function

$$\Xi_{group}(x, p, m) = \sum_{N=p}^{\infty} x^N Z_{group}(N, p, m). \quad (A.5)$$

In the grand-canonical picture, there are $p$ bound sequence-groups – each group is a set of $m$ bound sequences with lengths that fluctuate in unison – and 1 unbound sequence-group – a set of $m$ unbound sequences with lengths that fluctuate in unison. Each of the $p+1$ sequence-groups, fluctuate in length independently of one another against force $x$. Thus, we have

$$\Xi_{group}(x, p, m) = [U(x, m, 0)]^p U_F(x, m) \quad (A.6)$$

where $U(x,m,0) = \sum_{l=1}^{\infty} x^l u(l)$ is the grand partition function of a bound sequence, and

$U_F(x,m) = \sum_{l=0}^{\infty} x^l u_F(l)$ is the grand partition function of the unbound sequence.

In the Leontovich method, we obtain the fixed $N$ free energy of $Z_{group}$ by taking the free energy of $\Xi_{group}$ and subtracting the energy of interaction with the force represented by $x$, obtaining

$$Z_{group} = \frac{[U(\hat{x},m,0)]^p U_F(\hat{x},m)}{\hat{x}^N}, \quad (A.7)$$

where the fugacity $\hat{x}$ is chosen so that, for each replica, the thermal average of the total sequence length is $N$. This is done by making $\log Z_{group}$ stationary with respect to $\hat{x}$.

We generalize $U(x,m,0)$ to $U(x,m,a) = x + \sum_{l=2}^{\infty} x^l / l^{mc-a}$, and sum to obtain

$$U(x,m,a) = x + \tilde{A}^m [\phi(x, m \cdot c - a) - x] \quad (A.8)$$

where $\tilde{A} = A e^{g^{(1)}/RT}$ and $\phi(x,s) = \sum_{l=1}^{\infty} x^l / l^s$ is a special function known both as the polylogarithm function in applied mathematics, and as the ring function in polymer physics[43]. Useful properties of this function may be found in a paper by Truesdell [44]. We can also sum the geometric series in $U_F$, to obtain

$$U_F(x,m) = \frac{-e^{\frac{g_F}{RT}m}}{x - e^{\frac{g_F}{RT}m}}. \quad (A.9)$$

*(ii) Phase behavior and the inter-group overlap*

We now consider $n/m$ independent copies of the $m$-replica group discussed above, were n is a large integer and $m \in \{1, \text{K}, n\}$. The macrostate of the $\frac{n}{m}$-group system is determined by $Z_{group}$ and associated thermal-averages. While preserving the visual image of the structure of this system, we continue $m$ to the unit interval in all analytical expression for $Z_{group}$, the fugacity $\hat{x}$, and thermal-averages. We thus obtain the "continued" macrostate of the $\frac{n}{m}$-group system with the following properties.

We now calculate the overlap between replicas in distinct group when the macrostate of the $\frac{n}{m}$-group system has been continued. For a given $(\theta, m, T)$, and $F = 0$, the unbound sequence has finite (of order $N^0$) extent and does not contribute to the thermodynamics of the system; the unbound fraction is zero. Ignoring boundary effects, we assume that for each replica, the fraction of the time that a sequence-position is occupied by a monomer is independent of sequence-position. This assumption results in a thermal-average overlap of $q_u(\theta, m, T, F)\big|_{F=0} = \theta^2$.

### 3. "Minimizing" the variational free energy in the replica method

The stationarity equations result from zeroing the partial derivatives of $\beta f_0$, i.e.

$$t_\theta(\theta,m,T) \equiv m\frac{\partial \beta f_0}{\partial \theta} = \beta g^{(1)} m - \frac{\left[\beta g^{(2)}\right]^2}{2} m^2 [1-2\theta] - \log\left[U(\hat{x},m,0)\right]$$

$$t_m(\theta,m,T) \equiv \frac{m^2}{\theta}\frac{\partial \beta f_0}{\partial m} = -\frac{\left[\beta g^{(2)}\right]^2}{2} m^2 [1-\theta] - \left\{\frac{mU'(\hat{x},m,0)}{U(\hat{x},m,0)} - \log U(\hat{x},m,0)\right\} - \frac{1}{\theta}\log \hat{x}$$

(A.10)

where $\tilde{A} = Ae^{g^{(1)}/RT}$, $U'(x,m,a) = \partial U(x,m,a)/\partial m$, and

$$\theta = \frac{U(\hat{x},m,0)}{U(\hat{x},m,1)} \tag{A.11}$$

At large integer n, $m \in [1,n]$, so that n the limit $n \to 0$, $m \in (0,1]$ and minimization with respect to $m$ becomes maximization. At each value of $T$, the numerical search for the values of $\theta$ and $m$ that zeros Eq. (A.10) is performed in terms of $\hat{x}$ and $m$.

If we apply the uRSB scheme, see Section IV.B, to the homopolymer case $g^{(2)} = 0$ we obtain the single stationarity equation

$$t_\theta(\theta,m,T)\big|_{\substack{m=1 \\ g^{(2)}=0}} = \beta g^{(1)} - \log U(\hat{x},1,0) = 0. \tag{A.12}$$

### 4. Minimizing the variational free energy of the homogeneous-sequence model

For the DNA model described in Section II.A, we may obtain the thermodynamics of the homopolymer, $\delta g_t = 0$, by a route alternative to taking special limits of the replica procedure. By the maximum-term method, we can obtain an equation for the thermodynamically dominant value of the helix fraction. We start with the constrained partition function

$$Z_{\text{homo}}(p,T) = e^{-p\frac{g^{(1)}}{RT}} \sum_{l_1=1}^{\infty} \text{L} \sum_{l_{p=1}}^{\infty} \delta(N, \sum_{i=1}^{p} l_F) \prod_{i=1}^{p} u(l_i) \tag{A.13}$$

where the system is constrained to have $p$ helix-state bp's. As in Appendix A.2, the partition function of a bound sequence of length $\ell$ is

$$u(l) = \begin{cases} 1 & , l=1 \\ \dfrac{A}{-c} e^{\frac{g^{(1)}}{RT}} & , l \geq 2 \end{cases} \tag{A.14}$$

where the zero of energy is set to the free energy of the helix-state of a bp. In the maximum-term method [10, 45] the unconstrained partition function

$$Z_{\text{homo}}(T) = \sum_{p=0}^{N} Z_{\text{homo}}(p,T) \tag{A.15}$$

is approximated by the dominant value of $p$; this approximation becomes exact in the thermodynamic limit $N$ going to infinity.

The dominant value of $p$ is determined by the minimization equations

$$\begin{aligned}\frac{\partial}{\partial \theta} f_{\text{homo}}(\theta,T) &= 0 \\ \frac{\partial^2}{\partial \theta^2} f_{\text{homo}}(\theta,T) &> 0\end{aligned} \tag{A.16}$$

where

$$f_{\text{homo}}(\theta,T) = -\frac{RT}{N} \log Z_{\text{homo}}(p,T) \tag{A.17}$$

and $p = \lceil N\theta \rceil$, where $\lceil \cdots \rceil$ is the ceiling function. To obtain an analytical expression for $f_{\text{homo}}(\theta,T)$, we proceed as in Appendix A.1, and apply the Leontovich method, obtaining

$$Z_{\text{homo}}(\theta,T) = e^{-p\frac{g^{(1)}}{RT}} \frac{[U(\hat{x},1,0)]^p}{\hat{x}^N} \tag{A.18}$$

where $U(x,1,0)$, is the partition function of a bound sequence in contact with a bp reservoir at fugacity $x$, and is defined in Appendix A.2. After Eq. (A.18) is plugged into Eq. (A.17), the stationarity equation in Eq. (A.16) becomes

$$\frac{\partial}{\partial \theta} f_{\text{homo}}(\theta,T) = \beta g^{(1)} - \log U(\hat{x},1,0) = 0, \tag{A.19}$$

where $\theta = U(\hat{x},1,0)/U(\hat{x},1,1)$. Eq. (A.19) is the same as Eq. (A.12) in Appendix A.3.

58

# Figures

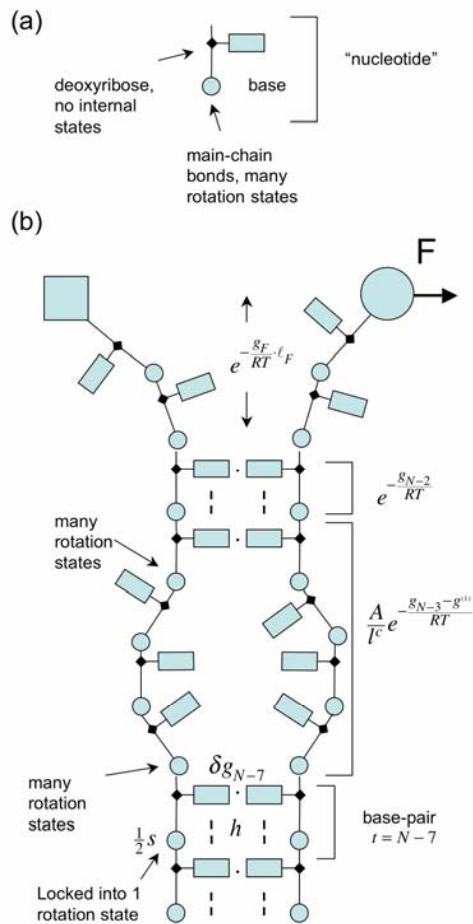

**Figure 1. Our model for dsDNA under unzipping force.** (a) We choose a partitioning of ssDNA into "nucleotides" that is convenient to describe unzipping. (b) Base-pair $t = N - 7$ is in the helix state, as is the preceding bp, so bp $N - 7$ contributes entropy $s$ due to the loss of rotation states of the main-chain bonds, enthalpy $h$ due to stacking with the previous bp, and free energy $\delta g_t$ due to hydrogen bonding. Bp $t = N - 7$ is followed by a bound sequence of length $\ell = 4$, which is followed by a bound sequence of length one, which is followed by the unbound sequence, having length $\ell_F = 2$. The large circle is the bead, which is pulled with constant force by the magnetic field. The big square represents the glass housing and is fixed in space.

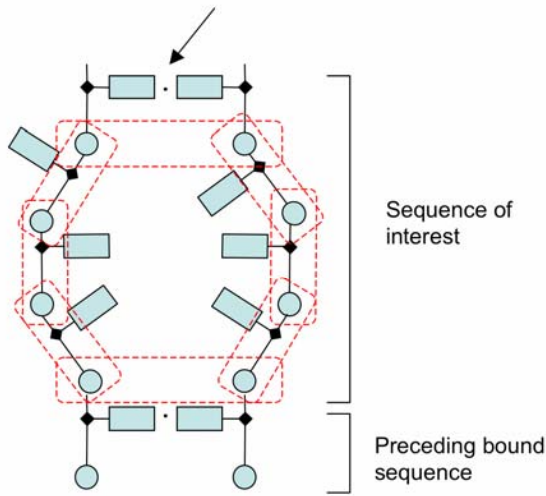

**Figure 2. Thermodynamics of a loop, i.e., a bound sequence with $\ell \geq 2$ bp's.** For the sequence of interest, $\ell = 4$. Here, we imagine that no bp's follow the sequence of interest. If the terminal bp in the sequence is removed from the helix-state (breaking the indicated h-bond), the nucleotides in both strands have the maximal number of rotation states; we call this the free coil macrostate. This macrostate has a free energy, purely entropic, that we set to be the zero of free energy (defined for $\ell \to \infty$). If the terminal bp is constrained to the helix-state, then the sequence of bp's forms a loop in which the individual monomers are indicated with boxes. The number of boxes is twice the number of bp's ($2\ell$) in the sequence of interest. As $\ell$ becomes large, the ring of boxes sample a number of configurations (relative to the number of configurations in the free coil state) that is approximated by the scaling law $A/\ell^c$. The T-independent prefactor $A$ is associated with the configuration-space volume of the helix-state (arrow) of the terminal bp.

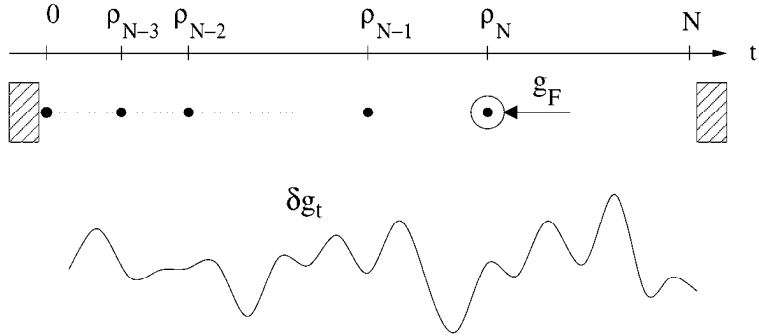

**Figure 3. 1-dimensional picture of our model.** Sequence-neighbors interact with pair potentials $u_A$ and $u_g$. Each monomer interacts with the external potential field $\delta g_t$.

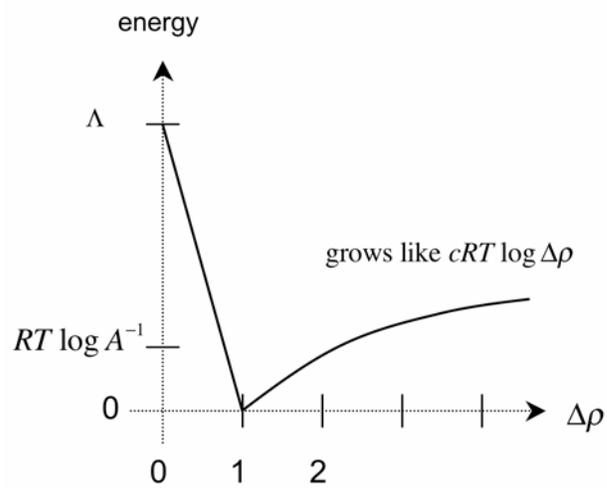

**Figure 4. In the 1-dimensional polymer, sequence connectivity can be represented as a pair-potential between sequence-neighbors.** Neighboring monomers cannot occupy the same sequence-position due to the large repulsive energy $\Lambda$. At large separations, a logarithmic attractive potential exists. This terms combine to give a pair potential with short-ranged repulsion and long-ranged attraction, and a narrow minimum at $\Delta\rho = 1$.

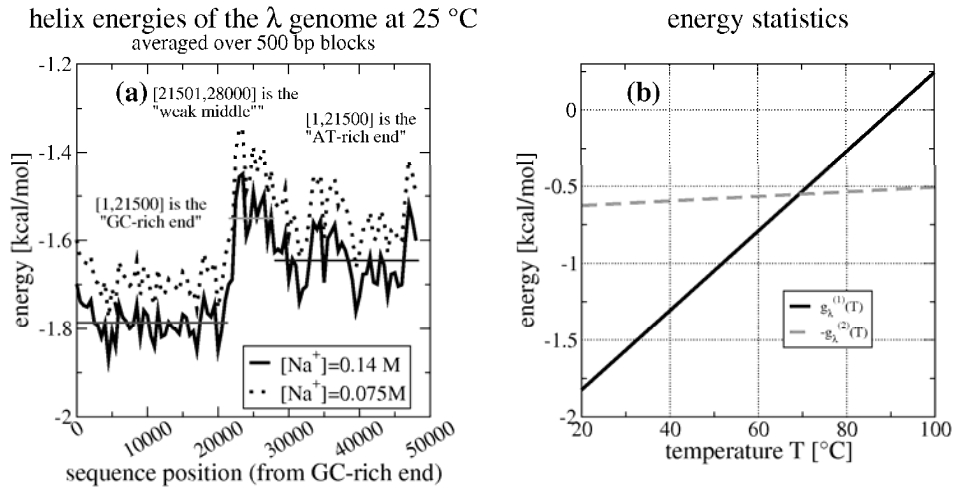

**Figure 5. Statistics of the NN free energies of the λ-phage genome.** Unless indicated otherwise, all calculations employ $[Na^+] = 0.14M$. (a) The NN energies are averaged over blocks of size 500 bp's (black curve). The average over the sub-sequence $[1, 21500]$ is $-1.75 kcal/mol$, the average over $[21501, 28000]$ is $-1.55 kcal/mol$, and the average over $[28001, 48501]$ is $-1.65 kcal/mol$ (horizontal lines). (b) The mean $g_\lambda^{(1)}$ and standard deviation $g_\lambda^{(2)}$, averaged over the whole genome, of the NN energies are shown at each temperature $T$. Note that we have plotted the negative of $g_\lambda^{(2)}$. For $T$ above $\approx 70°C$ the magnitude of $g_\lambda^{(2)}$ is comparable to or larger than the magnitude of $g_\lambda^{(1)}$. Thus, at high $T$, we expect heteropolymeric behavior.

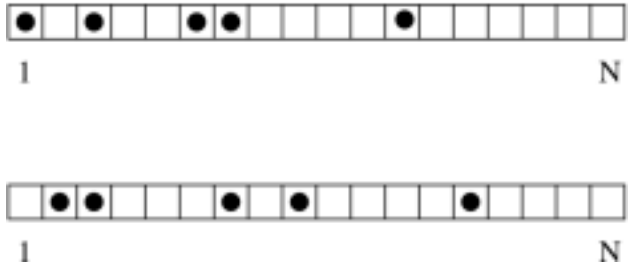

**Figure 6. Overlap between two replicas.** Each dot represents the presence of a monomer in the 1-dimensional polymer picture. The overlap is the fraction of positions at which both replicas exhibit a monomer.

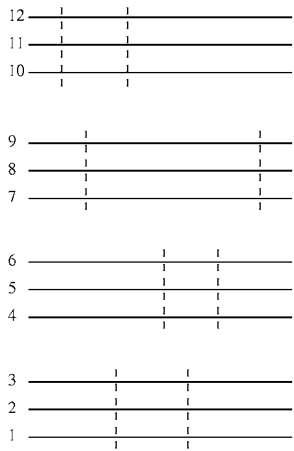

**Figure 7. m-braces do not span distinct groups.** Here, each group has size m=3. 2 m-braces enforce $p = 2$ overlaps between each pair of replicas in the same group. The overlap between a pair of replicas in distinct groups is the uncorrelated values $\theta^2$.

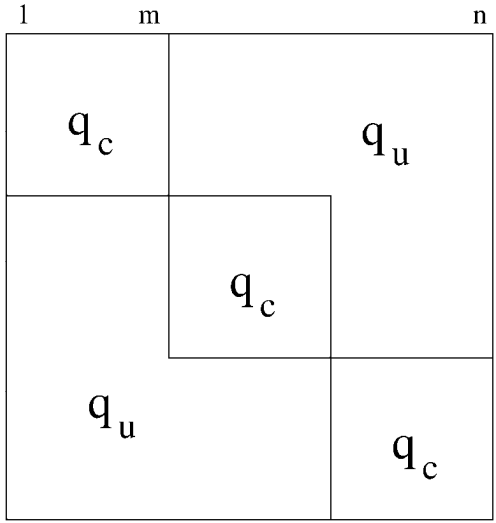

**Figure 8. The cuRSB overlap matrix.** In our implementation of the replica method, the replicas are divided into groups of size m. Within a group, the overlap between any two replicas has the maximally correlated value $q_c$, which is equal to the helix fraction, $q_c = \theta$. The groups fluctuate independently of each other, and the inter-group overlap has the value associated with the maximally uncorrelated value $q_u = \theta^2$.

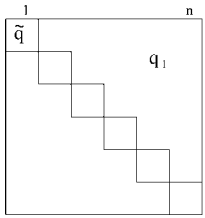

**Figure 9. The replica-symmetric matrix of overlaps.** The diagonal elements give the self-overlap, i.e., the helix fraction $\theta$. Each off-diagonal element give the overlap between a pair of replicas.

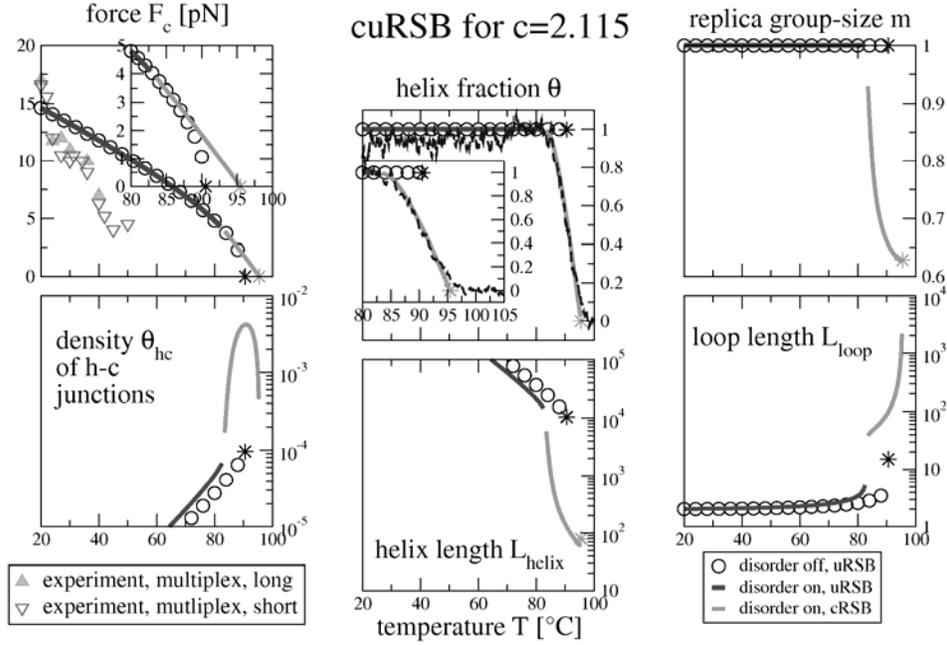

**Figure 10. The cuRSB scheme applied to the heteropolymer and homopolymer cases, for** $c = 2.115$. The black circles (homopolymer) and dark-grey line (heteropolymer) give the uRSB results, obtained by solving $\partial_\theta \beta f_0|_{m=1} = 0$ for $\theta$ and checking $\partial_m \beta f_0|_{m=1^-} > 0$, where $1^- = 1 - 10^{-6}$. The black star gives the solution $T_c^{homo}$ of Eq. **(29)**. We obtain $T_c^{homo} = 90.55\,°C$ for $c = 2.115$. The light-grey line (heteropolymer) give the cRSB results, obtained by solving $\partial_\theta \beta f_0 = 0$, $\partial_m \beta f_0 = 0$ for $\theta$, $m$, and checking that the stationary point is a local maximum in the $m$ direction. The light-grey star is the cRSB numerical proxy for $T_c$, obtained by solving $\partial_\theta \beta f_0 = 0$, $\partial_m \beta f_0 = 0$ for $T$, $m$ at the value of $\theta$ given by the right-hand-side of Eq. (A.11) for $\hat{x} = 1 - 10^{-12}$ ($\theta \cong 0$) and $m$ the solution of the stationarity equations; we obtain $T_c = 95.33\,°C$. Despite $c \geq 2$, the helix fraction falls continuously to zero, i.e., the melting transition is higher-order. We also show the data from single-molecule unzipping via magnetic tweezers; the up (down) triangles show the results using the protocol in which the sample is thermalized for a long (short) time [4]. In the plot of the helix fraction, we show an experimental circular dichroism (CD) melting curve (dashed line), which has been rescaled so that the low-T values fluctuate about 1. In the plots of the density of h-c junctions $\theta_{hc}$ and the helix length $L_{helix}$, the exponential dependence of the uRSB results (homopolymer and heteropolymer) continues down to $T = 20\,°C$.

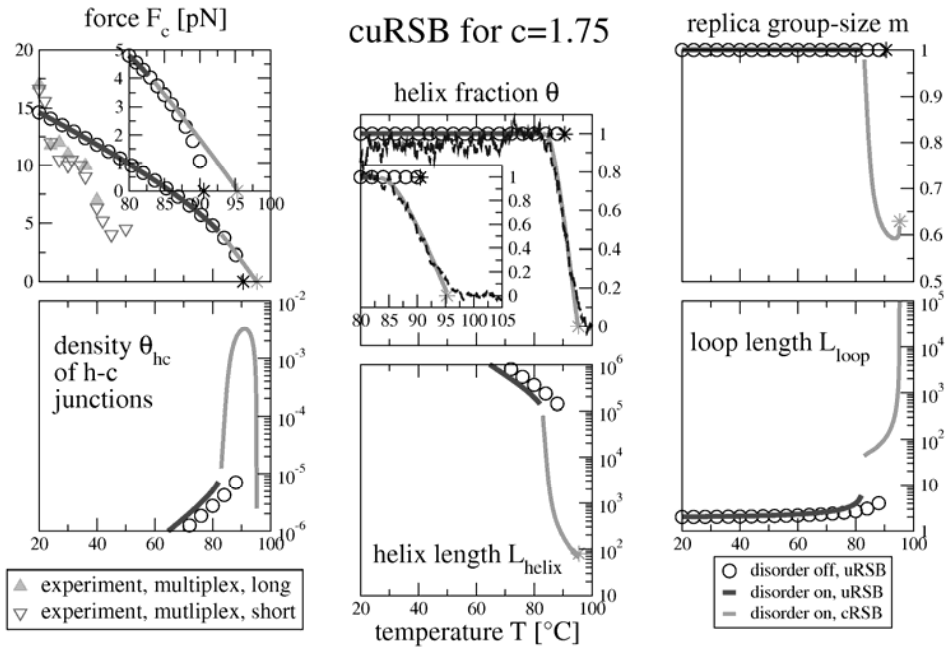

**Figure 11. The cuRSB scheme applied to the heteropolymer and homopolymer cases, for** $c = 1.75$. The data is the labeled and obtained in the same way as Figure 10. For these parameters, the random-sequence model has $T_c = 95.14\,°C$, whereas the homogeneous model has $T_c^{homo} = 90.55\,°C$, as for $c = 2.115$.

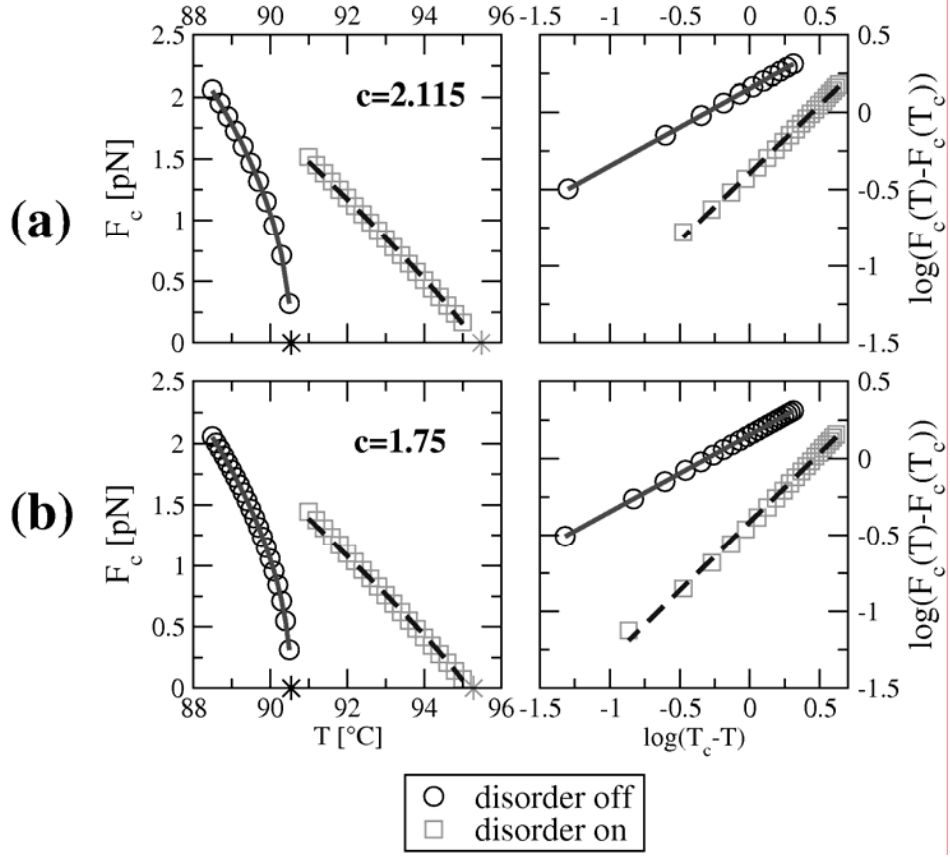

**Figure 12. Scaling law fit obtains an effective loop exponent $c^{fit}$, for (a) $c = 2.115$ and (b) $c = 1.75$.** The critical force curves are obtained from the cuRSB approximation scheme. We fit the curves to the scaling law in Eq. **(33)**, with $\alpha = 1/2\eta$, **Error! Reference source not found.** in order to obtain the effective loop exponent $c^{fit}$ for the random heteropolymer and homopolymer models. These exponents are obtained from the slopes of the straight-line fits in log-log scale, and the quality of the fit is measured in terms of $r^2$, the square of the Pearson correlation coefficient. **(a)** For $c = 2.115$, the heteropolymer is fit on $91 \leq T \leq 95\,°C$ giving $\alpha = 0.881$ with $r^2 = 0.998$, thus $c^{ef} = 1.57$. **(b)** For $c = 1.75$, the heteropolymer is fit on $91 \leq T \leq 95\,°C$ giving $\alpha = 0.899$ with $r^2 = 0.997$, thus $c^{ef} = 1.56$. For both $c = 2.115,\ 1.75$, the homopolymer is fit on $88.5 \leq T \leq 90.5\,°C$ and has exponent 0.502 with $r^2 = 0.999997$ that is consistent with Eq. **(32)**.

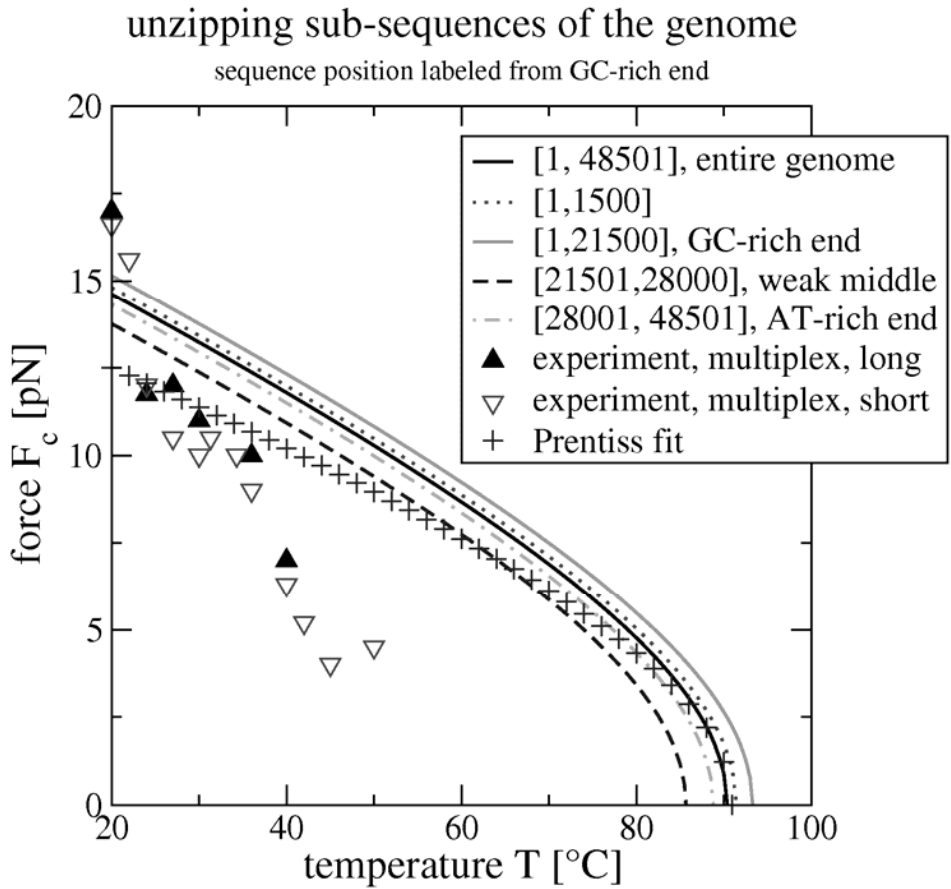

**Figure 13. The all-or-none model compared with the experimental unzipping of the 1st 1500 bp's.** For each curve labeled with a sub-sequence of the genome of λ-phage, we computed the critical force of an all-or-none model, with the free energy per bp of the all-helix state trained on the indicated sub-sequence. For comparison, we show the published results (triangles) of the multiplexed single-molecule experiment on the unzipping of the sub-sequence [1,1500]; this data is reproduced with permission [4]. We also show the critical force of the all-or-none model using the all-helix state free energy per bp given in [4].

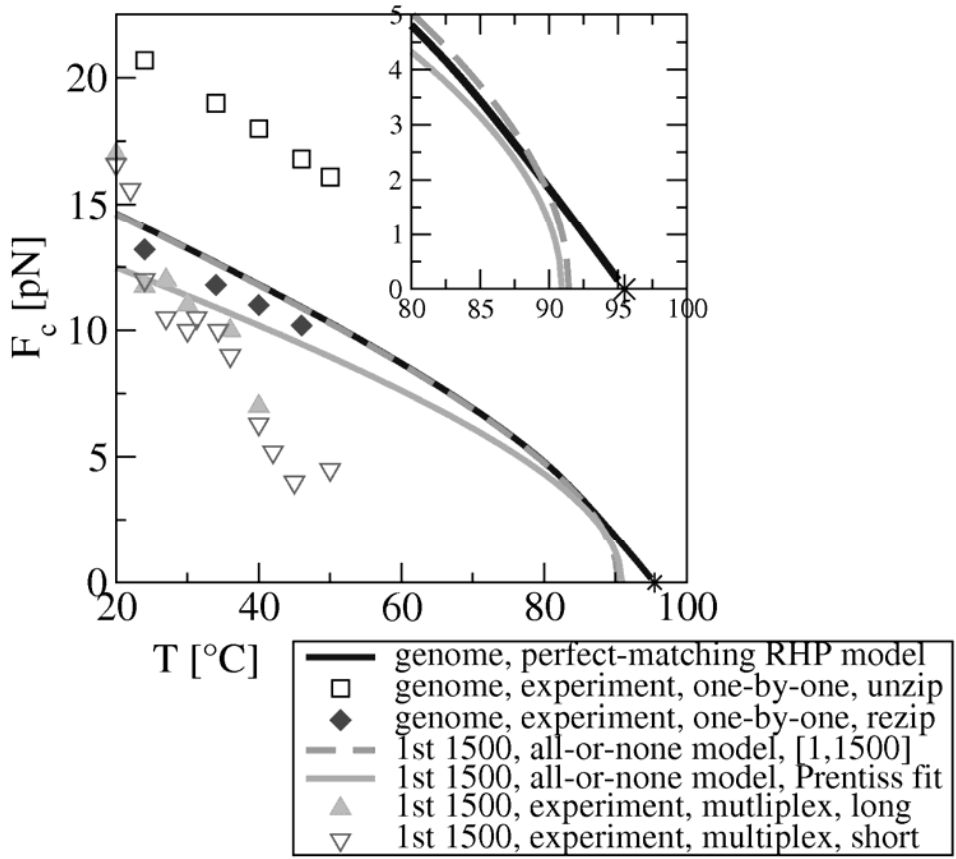

**Figure 14. The perfect-matching RHP model compared with the experimental unzipping of the entire genome.** In the first three data sets (black line, open squares, solid diamonds), we compare our perfect-matching RHP model to the results of experiments in which an individual copy of the phage-$\lambda$ genome is unzipped and then re-zipped by increasing and then decreasing the applied force at a constant rate (data unpublished). The second four data sets (dashed line, grey line, up triangle, down triangle) show how the all-or-none model compares with the experiment in which follows the unzipping of the first 1500 bp's of many copies of the genome at the same time (multiplex). The designations "long" and "short" refer to the duration of time at which the sample is allowed to thermalize at the target temperature.